\pdfoutput=1
\documentclass[a4paper,fleqn,usenatbib]{mnras}
\usepackage[T1]{fontenc}
\usepackage{ae,aecompl}
\usepackage{graphicx}	
\usepackage{amsmath}	
\usepackage{amssymb}
\usepackage{color}
\definecolor{notecolor}{rgb}{0.8,0,0}

\def\lya{Ly$\alpha$~} 

\title[21~cm models]{Models of the Cosmological 21~cm Signal from the
  Epoch of Reionization Calibrated with \lya and CMB Data}

\author[Kulkarni et al.]{{Girish Kulkarni$^{1}$\thanks{Email:
      kulkarni@ast.cam.ac.uk}, Tirthankar Roy Choudhury$^{2}$, Ewald
    Puchwein$^1$} \newauthor{and Martin G. Haehnelt$^1$} \\ $^1$Institute
  of Astronomy and Kavli Institute of Cosmology, University of
  Cambridge, Madingley Road, Cambridge CB3 0HA, UK \\ $^2$National
  Centre for Radio Astrophysics, Tata Institute of Fundamental
  Research, Pune 411007, India}

\date{Accepted ---. Received ---; in original form ---}

\pubyear{2016}

\begin{document}
\label{firstpage}
\pagerange{\pageref{firstpage}--\pageref{lastpage}}
\maketitle

\begin{abstract}
  We present here 21~cm predictions from high dynamic range
  simulations for a range of reionization histories that have been
  tested against available \lya and CMB data.  We assess the
  observability of the predicted spatial 21~cm fluctuations by ongoing
  and upcoming experiments in the late stages of reionization in the
  limit in which the hydrogen spin temperature is significantly larger
  than the CMB temperature.  Models consistent with the available \lya
  data and CMB measurement of the Thomson optical depth predict
  typical values of 10--20 mK$^2$ for the variance of the 21~cm
  brightness temperature at redshifts $z=7$--$10$ at scales accessible
  to ongoing and upcoming experiments ($k \lesssim 1$ cMpc$^{-1}h$).
  This is within a factor of a few magnitude of the sensitivity
  claimed to have been already reached by ongoing experiments in the
  signal rms value.  Our different models for the reionization history
  make markedly different predictions for the redshift evolution and
  thus frequency dependence of the 21~cm power spectrum and should be
  easily discernible by LOFAR (and later HERA and SKA1) at their
  design sensitivity.  Our simulations have sufficient resolution to
  assess the effect of high-density Lyman limit systems that can
  self-shield against ionizing radiation and stay 21~cm bright even if
  the hydrogen in their surroundings is highly ionized. Our
  simulations predict that including the effect of the self-shielded
  gas in highly ionized regions reduces the large scale 21~cm power by
  about 30\%.
\end{abstract}

\begin{keywords}
dark ages, reionization, first stars -- galaxies: high-redshift -- intergalactic medium
\end{keywords}

\section{Introduction}

Finally unraveling the complete ionization history of hydrogen with
high-redshift 21~cm observations is the major science driver of
currently operating and planned low-frequency radio telescopes.
Achieving the necessary dynamic range for accurate models of the
reionization process has thereby been recognized as a key challenge
\citep{2011ASL.....4..228T}.  Numerical simulations that aim to be
self-consistent in their modeling of ionizing sources and the
radiative transfer of ionizing photons in the intergalactic medium
(IGM) cannot account for the clustering of sources or the structure of
the ionization field on scales greater than $\sim 10$ comoving Mpc
\citep{2012MNRAS.427.2464F, 2013MNRAS.436.1818F, 2014ApJ...789..149S,
  2015MNRAS.451.1586P}.  On the other hand, simulations that can take
these large scale effects into consideration have low spatial and mass
resolution and are unable to consistently model small-scale effects
such as radiative feedback on ionizing sources and self-shielding of
high density regions in the IGM \citep{2007ApJ...659..865L,
  2009MNRAS.393...32T, 2010ApJ...724..244A, 2012ApJ...756L..16A,
  2012AIPC.1480..248S, 2015MNRAS.453.3593B, 2015MNRAS.454.1012A}.

Simulations are nevertheless crucial for the ongoing and upcoming
experiments that aim to detect the 21~cm signal from the epoch of
reionization \citep{2015aska.confE...7I}.  The 21~cm brightness
distribution is expected to eventually become the ultimate probe of
reionization.  The design of instruments capable of detecting this
signal is guided by its predictions from numerical simulations
\citep[e.g.,][]{2012ApJ...753...81P}.  Simulations are also crucial in
the interpretation of the results of these experiments
\citep{2016MNRAS.455.4295G}, all of which aim to detect the 21~cm
signal statistically.  Thus, given their relevance, not only should
these simulations have a large enough dynamic range to be
self-consistent and convergent but they should also be consistent with
other currently available constraints on the epoch of reionization.
Due to their computational cost, most simulations of the 21~cm signal
lack one or both of these properties.

It has been argued that the goal of self-consistent large scale
simulation of cosmic reionization is now gradually coming within reach
thanks to Moore's Law \citep{2014ApJ...793...29G, 2014ApJ...793...30G,
  2015arXiv151100011O, 2015ApJS..216...16N}, but semi-numerical and
analytical methods of reionization modeling continue to remain
attractive for efficient and flexible exploration of the parameter
space, especially given the paucity of data at high redshifts
\citep{2007ApJ...669..663M, 2008MNRAS.386.1683G, 2009MNRAS.394..960C,
  2011MNRAS.411..955M, 2011MNRAS.412.2781K, 2011MNRAS.417.2264V,
  2012MNRAS.423..862K, 2012ApJ...747..126A, 2013MNRAS.428L...1M,
  2013RAA....13..373Z, 2013ApJ...776...81B, 2013ApJ...768...71R,
  2013ApJ...771...35K, 2014MNRAS.440.1662S, 2014MNRAS.442.1470P,
  2015MNRAS.454L..76M, 2016MNRAS.457.1550H}.

In this paper, we combine a high dynamic range cosmological simulation
with an excursion set based model for the growth of ionized regions to
predict the 21~cm signal during the epoch of reionization.  We follow
the approach of \citeauthor{2015MNRAS.452..261C}
(\citeyear{2015MNRAS.452..261C}; hereafter CPHB15) to calibrate the
simulation parameters such that they reproduce the IGM Lyman-$\alpha$
(Ly$\alpha$) opacity at $z \lesssim 6$ \citep{2006ARA&A..44..415F,
  2015MNRAS.447.3402B, 2015MNRAS.447..499M} as well as the cosmic
microwave background (CMB) constraints on the electron scattering
optical depth \citep{2016arXiv160502985P, 2016arXiv160503507P}.  The
advantage of this approach is that once the reionization history is
given, all other quantities of interest---such as the photoionization
rate, emissivity of ionizing sources, and the clumping factor of the
IGM---can be calculated self-consistently from the simulation box at
each redshift.  This enables us to simulate concordant models of
reionization consistent with a wide variety of observations without
losing the dynamic range of our simulations.

CPHB15 applied this method to study the evolution of Ly$\alpha$
emission in high-redshift galaxies by calibrating a ``hybrid''
cosmological simulation box, which was created by combining a
low-resolution cosmological simulation box at large scales with a
high-resolution simulation box at small scales.  A similar approach
was used by \citet{2015MNRAS.446..566M} to study the evolution of the
Lyman-$\alpha$ emitter fraction of high-redshift galaxies.  In their
approach, a seminumerical scheme was used to obtain the low-resolution
simulation. The hybrid box used in CPHB15 formally had very high
dynamic range (equivalent to a cosmological simulation with $2\times
5120^3$ particles in a 100 $h^{-1}$cMpc box) but did not correctly
represent the clustering of matter at scales larger than the size of
the small box, which was 10 $h^{-1}$cMpc.  Thus from the point of view
of deriving the cosmological 21~cm signal, hybrid boxes are of little
use as they fail to yield, e.g., the 21~cm power spectrum at scales of
interest.  The main improvement in the simulation method used in this
work is the use of a cosmological hydrodynamical simulation with
improved dynamic range.

We follow CPHB15 and consider three different reionization histories
for our analysis.  One of our reionization histories follows the
widely used model of the meta-galactic UV background by
\citeauthor{2012ApJ...746..125H} (\citeyear{2012ApJ...746..125H};
hereafter HM12).  This model was tuned to match the constraints on
reionization from the Wilkinson Microwave Anisotropy Probe (WMAP;
\citealt{2011ApJS..192...14J}) and predicts an electron scattering
optical depth higher than the recent Planck measurements
\citep{2016arXiv160502985P, 2016arXiv160503507P}.  We therefore
explore two other reionization histories in which reionization is
completed later than in the HM12 model and the electron scattering
predictions are consistent with Planck results.

In addition to the evolution of the ionized fraction, the 21~cm signal
also depends on the distribution of optically thick systems that are
self-shielded from the ionizing radiation.  Such systems, which are
high-redshift counterparts of the Lyman-limit systems seen in quasar
absorption spectra, are usually missed by low resolution simulations.
We leverage our high dynamic range to study the effect of these
self-shielded regions on the 21~cm signal using a prescription for
self-shielding provided by \citet{2013MNRAS.430.2427R}.  Finally, we
consider whether our predicted signal can be observed by five ongoing
and upcoming 21~cm experiments.  The main aim of the paper is thus to
use models that are calibrated to existing data and predict the 21~cm
signal and its detectability at different redshifts.

We describe our simulations and the calibration procedure in
Section~\ref{sec:sims}.  Section~\ref{sec:21cm} presents our
predictions for the 21~cm signal and its observability in ongoing and
future experiments.  We investigate the effect of various assumptions
on our results in Section~\ref{sec:ss_effect} and conclude with a
discussion in Section~\ref{sec:end}.  Our $\Lambda$CDM cosmological
model has $\Omega_\mathrm{b}=0.0482$, $\Omega_\mathrm{m}=0.308$,
$\Omega_\Lambda=0.692$, $h=0.678$, $n=0.961$, $\sigma_8=0.829$, and
$Y_\mathrm{He}=0.24$ \citep{2014A&A...571A..16P}.

\begin{figure*}
\begin{center}
  \begin{tabular}{cc}
    \includegraphics*[width=\columnwidth]{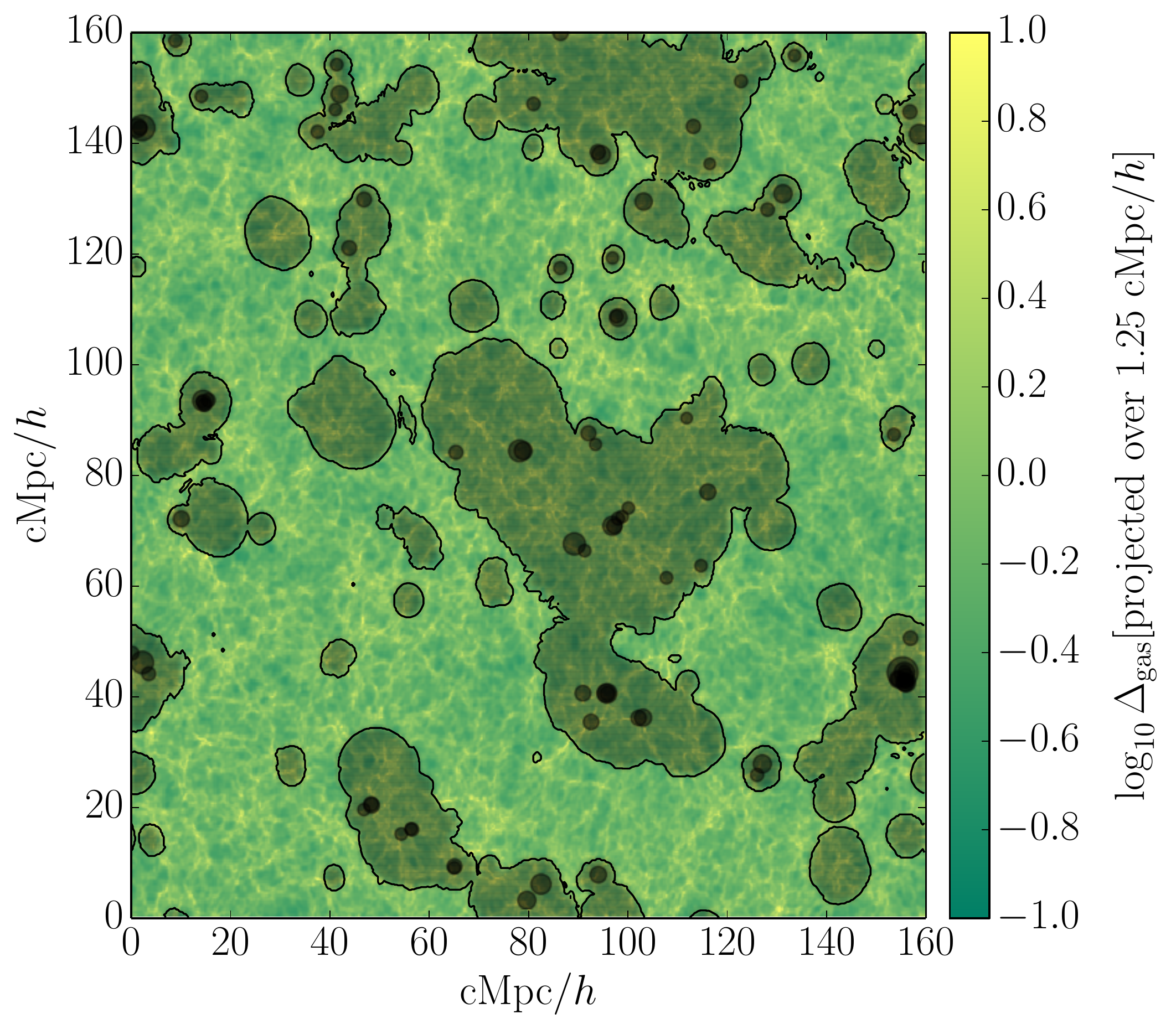} &
    \includegraphics*[width=\columnwidth]{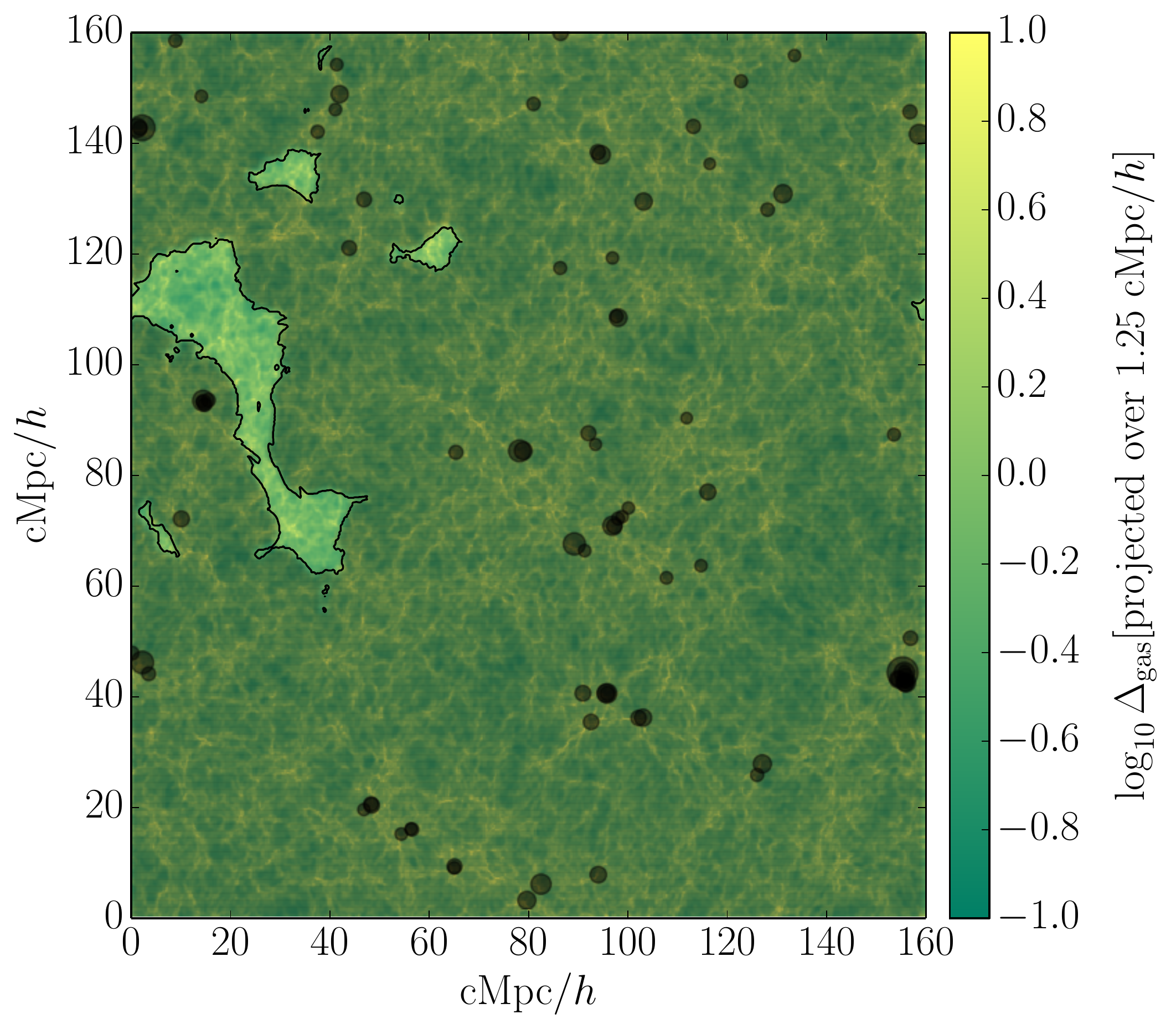}
  \end{tabular}
\end{center}
\caption{Distribution of gas density at $z=7$.  Black symbols denote
  the locations of centres of masses of dark matter haloes. Shaded
  areas in both panels show ionized regions identified by the
  excursion set method.  Colour scale shows the gas density.  The left
  panel shows the ionization field for $\zeta_\mathrm{eff}=0.5$, which
  corresponds to $Q_V=0.31$.  The right panel shows the ionization
  field for $\zeta_\mathrm{eff}=1.0$, which corresponds to
  $Q_V=0.94$.}
\label{fig:slices}
\end{figure*}

\section{Calibrated Simulations of Cosmic Reionization}
\label{sec:sims}

Our 21 cm predictions are based on cosmological hydrodynamical
simulations with large dynamic range that are part of the Sherwood
simulation suite and were run as part of a large (15 million core
hour, PI: James Bolton) \textsc{prace} simulation program
\citep{2016arXiv160503462B}. Sources of ionizing radiation are placed
in dark matter haloes identified in the simulation, and an ionization
field is obtained using the well-known excursion set approach.  This
resultant ionization field is then calibrated to a given reionization
history, while accounting for residual neutral gas in ionized regions.
The reionization history used for calibration is chosen carefully such
that it is consistent with a range of \lya and CMB data.  In this
manner, our models self-consistently predict the large scale
distribution of neutral hydrogen with high resolution for reionization
histories consistent with data that constrain the ionization state of
hydrogen during the late stages of reionization.

\subsection{Large Scale Ionization Field}

The Sherwood simulation suite has been run using the energy- and
entropy-conserving TreePM smoothed particle hydrodynamical (SPH) code
\textsc{p-gadget-3}, which is an updated version of the
\textsc{gadget-2} code \citep{2001NewA....6...79S,
  2005MNRAS.364.1105S}.  Our base simulation was performed in a cubic
box of length 160 $h^{-1}$cMpc on a side. Periodic boundary conditions
were imposed.  The number of gas and dark matter particles were both
initially $2048^3$.  This corresponds to a dark matter particle mass
of $M_\mathrm{dm}=3.44\times 10^7$ $h^{-1}$M$_\odot$ and gas particle
mass of $M_\mathrm{gas}=6.38\times 10^6$ $h^{-1}$M$_\odot$.  The
softening length was set to $l_\mathrm{soft}=3.13$ $h^{-1}$ckpc.  The
simulation evolves the gas and dark matter density fields from $z=99$
to $z=2$.  We use the {\tt QUICK\_LYALPHA} flag in
\mbox{\textsc{p-gadget-3}} in order to speed up the simulation: gas
particles with temperature less than $10^5$~K and overdensity of more
than a thousand times the mean baryon density are converted to
collisionless stars and removed from the hydrodynamical calculation
\citep{2004MNRAS.354..684V}.

In addition to the cosmological evolution of baryons and dark matter,
the simulation implements photoionization and photoheating of baryons
by calculating the equilibrium ionization balance of hydrogen and
helium in a optically thin UV background based on the model of
\citet{2012ApJ...746..125H}, modified so that the resultant IGM
temperature agrees with the measurements by
\citet{2011MNRAS.410.1096B}.  Radiative cooling is implemented by
taking into account cooling via two-body processes such as collisional
excitation of H~\textsc{i}, He~\textsc{i}, and He~\textsc{ii},
collisional ionization of H~\textsc{i}, He~\textsc{i}, and
He~\textsc{ii}, recombination, and Bremsstrahlung
\citep{1996ApJS..105...19K}.  Likewise, \textsc{p-gadget-3} also
includes inverse Compton cooling off the CMB
\citep{1986ApJ...301..522I}, which can be an important source of
cooling at high redshifts. We ignore metal enrichment and its effect
on cooling rates, which is a good approximation for the IGM.  In the
redshift range relevant to this paper, we use snapshots of the
particle positions at $z=10, 8, 7,$ and $6$.  Dark matter haloes are
identified using the friends-of-friends algorithm.  To calculate power
spectra, we project the relevant particles onto a grid to create a
density field, using the cloud-in-cell (CIC) scheme.  After
calculating the power spectrum, we deconvolve the CIC kernel, ignoring
small errors due to aliasing on the smallest scales
\citep{2008ApJ...687..738C}.

Having obtained the gas density field from the cosmological
simulation, we then derive the ionization field corresponding to a
distribution of sources with some ionizing emissivity.  We assume that
the total number of ionizing photons produced by a halo, $N_\gamma$,
is proportional to its mass $M$.  (We further discuss and vary this
assumption in Section~\ref{sec:nonlin}.)  The minimum halo mass in our
simulation is $1.6\times 10^{8}$~$h^{-1}$M$_\odot$; the maximum halo
mass at $z=7$ is $2.1\times 10^{12}$~$h^{-1}$M$_\odot$.  A grid cell
at position $\mathbf{x}$ is ionized if the condition
\begin{equation}
  \langle n_\gamma(\mathbf{x})\rangle_R > \langle n_\mathrm{H}(\mathbf{x})\rangle_R(1+\bar N_\mathrm{rec}),
  \label{eqn:exset1}
\end{equation}
is satisfied in a spherical region centred on the cell for some radius
$R$ \citep{2004ApJ...613....1F, 2009MNRAS.394..960C,
  2011MNRAS.411..955M}.  Here, the averages are over the spherical
region, $n_\mathrm{H}$ is the hydrogen number density, 
\begin{equation}
  n_\gamma = \int_{M_\mathrm{min}}^\infty dM\frac{dn}{dM}N_\gamma(M),
  \label{eqn:ngamma}
\end{equation}
where $dn/dM$ is the halo mass function within the spherical region,
$M_\mathrm{min}$ is the minimum halo mass that contributes ionizing
photons, $N_\gamma(M)$ is the number of ionizing photons from a halo
of mass $M$, and $\bar N_\mathrm{rec}$ is the average number of
recombinations per hydrogen atom in the IGM.  The condition in
Equation~(\ref{eqn:exset1}) can be recast as
\begin{equation}
  \zeta_\mathrm{eff}f(\mathbf{x},R)\geq 1,
  \label{eqn:exset}
\end{equation}
where
\begin{equation}
  f\propto \rho_m(R)^{-1}\int_{M_\mathrm{min}}^\infty dM\frac{dn}{dM}N_\gamma(M).
\end{equation}
Here $\rho_m(R)$ is the average matter density in the sphere of radius
$R$.  The quantity $f$ is identical to the collapsed fraction
$f_\mathrm{coll}$ if $N_\gamma(M)\propto M$.  The parameter
$\zeta_\mathrm{eff}$ here is the effective ionizing efficiency, which
corresponds to the number of photons in the IGM per hydrogen atom in
stars, compensated for the number of hydrogen recombinations in the
IGM.  It is the only parameter that determines the large scale
ionization field in this approach.  Cells that do not satisfy the
criterion in Equation~(\ref{eqn:exset}) are at least partially
neutral, and are assigned an ionized fraction
$\zeta_\mathrm{eff}f(\mathbf{x},R_\mathrm{min})$, where
$R_\mathrm{min}$ is the length of the cell.  We denote the ionized
volume fraction in a cell $i$ as $Q_i$.  The total volume-weighted
ionized fraction is then $Q_V\equiv \sum_iQ_i/n_\mathrm{cell}$, where
$n_\mathrm{cell}$ is the number of grid cells.

Figure~\ref{fig:slices} shows the ionization field in a
1.25~$h^{-1}$cMpc deep slice of our simulation.  The colour scale
shows the gas density distribution.  Black symbols denote the
locations of the centres of masses of dark matter haloes.  The left and
right panels of Figure~\ref{fig:slices} show the ionization field
corresponding to $\zeta_\mathrm{eff}=0.5$ and $1$, respectively.  As
expected, the volume-weighted ionization fraction increases with
$\zeta_\mathrm{eff}$.  Ionized regions prefer overdensities around
dark matter haloes.  Similar large-scale ionization fields obtained
using the excursion set approach have been shown to agree with results
of low resolution radiative transfer simulations
\citep{2014MNRAS.443.2843M}.  However, the excursion set method misses
high resolution features, such as self-shielded high density sinks of
ionizing photons within the ionized regions
\citep{2014MNRAS.440.1662S}.  A second, related, drawback is that the
distribution of ionized IGM in Figure~\ref{fig:slices} is not
calibrated to any observational constraints.

\begin{figure}
  \begin{center}
    \includegraphics[width=\columnwidth]{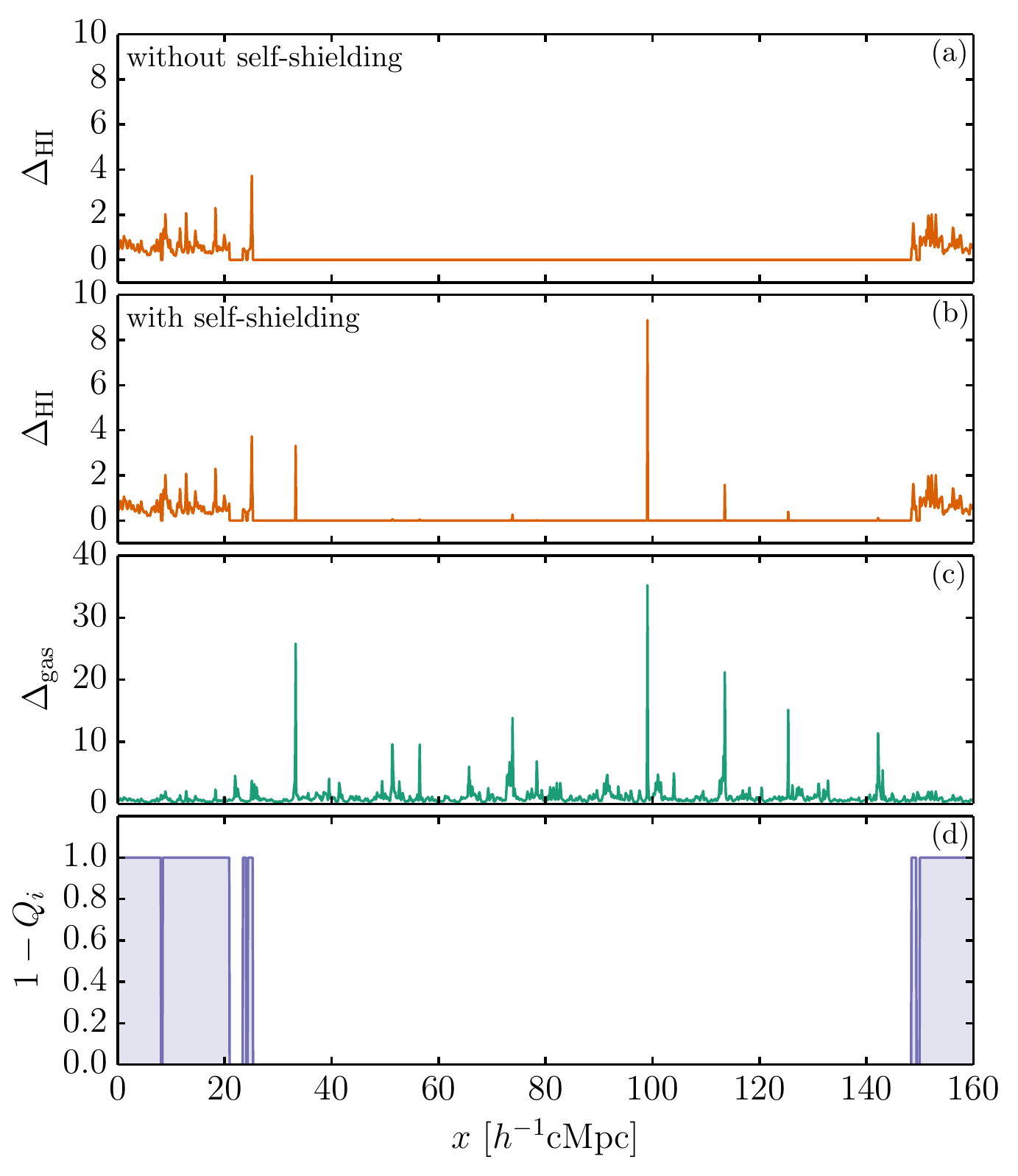}
  \end{center}
  \caption{Neutral hydrogen overdensity ($\Delta_\mathrm{HI}\equiv
    n_\mathrm{HI}/\bar n_\mathrm{H}$) along a randomly selected
    sightline at $z=7$ from the Late/Default model.  Panel (a) shows
    the neutral hydrogen density distribution in absence of
    self-shielding; panel (b) shows the density distribution when
    self-shielding is assumed.  Panels (c) and (d) show the total gas
    overdensity ($\Delta_\mathrm{gas}\equiv n_\mathrm{gas}/\bar
    n_\mathrm{gas}$) and the large scale ionization field (here shown
    as the neutral hydrogen fraction, $1-Q_i$) along the same
    sightline.  The region from about $30$ to $150$ $h^{-1}$cMpc is
    ionized, but high density locations within this region can
    self-shield.}
  \label{fig:skewers}
\end{figure}
  
\subsection{Calibration}
\label{sec:calibration}

We calibrate the large scale ionization field obtained by the
procedure described above to a  chosen reionization history,
incorporating inhomogeneities within ionized regions.  This is done
using the method developed by CPHB15, which we now describe.  The aim
here is to use the density field from the hydrodynamical simulation
and the ionization field from the excursion set approach to derive the
spatial distribution of the photoionization rate $\Gamma_\mathrm{HI}$
that reproduces a given reionization history.  As we will see below,
this calibration is equivalent to solving the globally averaged
radiative transfer equation at high resolution, without the
concomitant numerical cost.

\begin{table*}
  \begin{center}
    \begin{tabular}{clccccccc}
      \hline
      & Model & $L_\mathrm{box}$ & $N_\mathrm{gas}$ & $M_\mathrm{gas}$ & $z_\mathrm{reion}$ & $\tau$ & Minimum halo mass & $N_\gamma(M)$ \\
      & & ($h^{-1}$cMpc) & & ($h^{-1}$M$_\odot$) & & & (M$_\odot$) & \\
      \hline
      1. & HM12 & 160 & $2048^3$ & $6.38\times 10^6$ & 6.7 & 0.084 & $2.3\times 10^8$ & $\propto M_\mathrm{halo}$ \\
      2. & Late/Default & 160 & $2048^3$ & $6.38\times 10^6$ & 6.0 & 0.068 & $2.3\times 10^8$ & $\propto M_\mathrm{halo}$ \\
      3. & Very Late & 160 & $2048^3$ & $6.38\times 10^6$ & 6.0 & 0.055 & $2.3\times 10^8$ & $\propto M_\mathrm{halo}$ \\
      \\
      2a. & High Mass & 160 & $2048^3$ & $6.38\times 10^6$ & 6.0 & 0.068 & $3.5\times 10^{10}$ & $\propto M_\mathrm{halo}$ \\
      2b. & Nonlinear & 160 & $2048^3$ & $6.38\times 10^6$ & 6.0 & 0.068 & $2.3\times 10^8$ & $\propto M_\mathrm{halo}^{1.41}$ \\
      2c. & Convergence Run & 40 & $2048^3$ & $9.97\times 10^4$ & 6.0 & 0.068 & $2.3\times 10^8$ & $\propto M_\mathrm{halo}$ \\
      \hline
    \end{tabular}
  \end{center}
  \caption{Reionization models considered in this paper.  Models 1--3
    represent the three reionization histories considered.  The
    ``Late/Default'' model is our preferred model of reionization.
    Models 2a and 2b are variations on the Late/Default model.  The
    reionization history in these models is identical to the
    Late/Default model, but other details have changed. The ``High
    Mass'' model has only relatively high mass haloes, while in the
    ``Nonlinear'' model, the number of ionizing photons contributed by
    a halo, $N_\gamma(M)$, defined in Equation~(\ref{eqn:ngamma}), has
    a nonlinear dependence on the halo mass.  Model 2c uses a higher
    resolution base simulation together with Late/Default reionization
    history to check our results for convergence; this model is
    discussed in Appendix~\ref{sec:conv}.}
  \label{tab:models}
\end{table*}

\begin{figure}
  \begin{center}
    \includegraphics[width=\columnwidth]{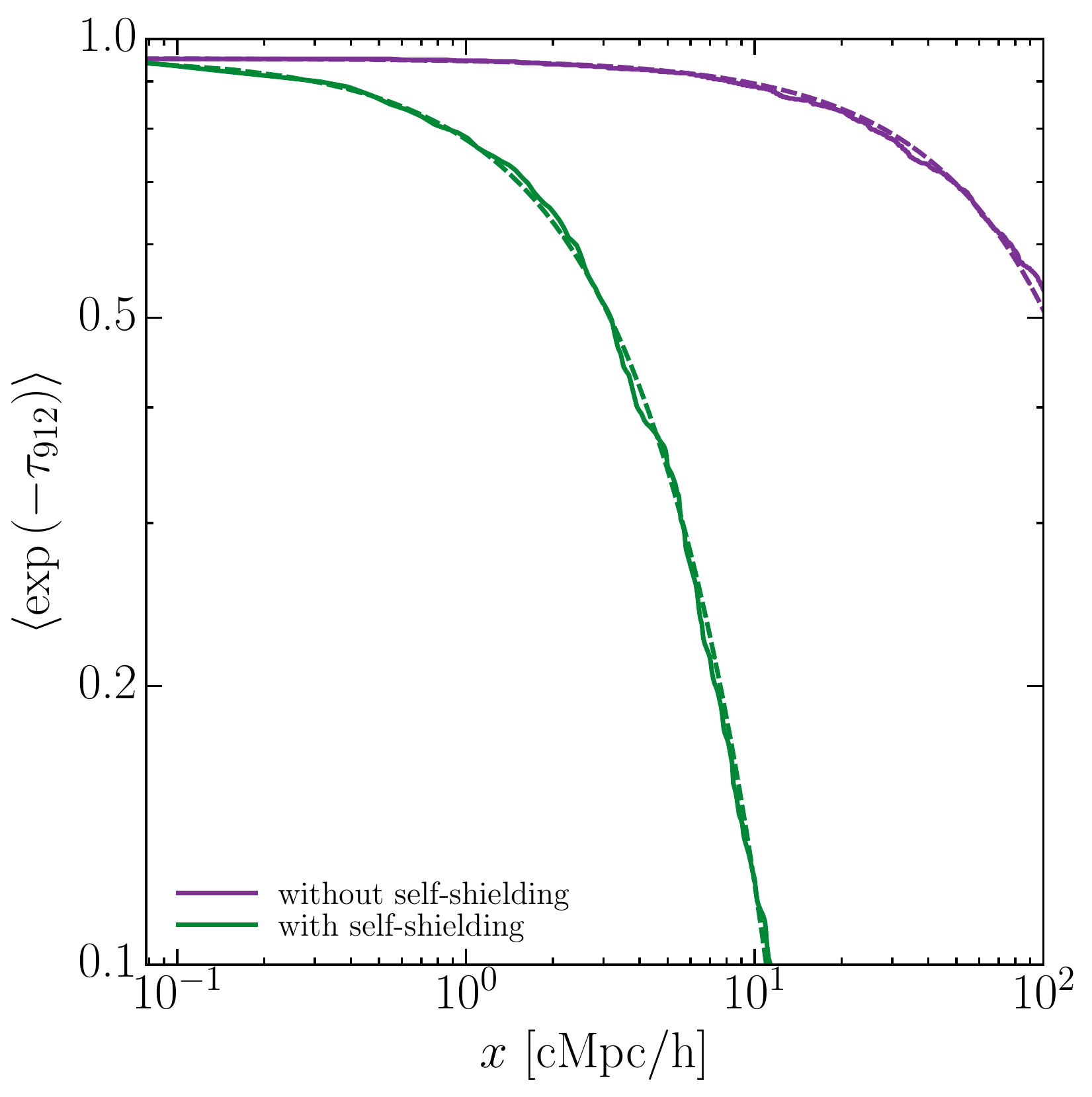}
  \end{center}
  \caption{The mean 912~{\AA} transmission along 1000 sightlines in
    the Late/Default model with and without self-shielding (green and
    purple curves, respectively).  Dashed curves show fits of the form
    given in Equation~(\ref{eqn:mfp}).}
  \label{fig:flux}
\end{figure}

Following CPHB15, we consider three reionization histories for
calibration:

\begin{itemize}

\item {\bf HM12:} This model corresponds to the ionization history 
  predicted by the widely used  model of the meta-galactic UV background by 
   \citet{2012ApJ...746..125H}.  In this model,
  ionized regions overlap and the universe is completely reionized at
  $z=6.7$.  The galaxy UV emissivity traces the cosmic star formation
  history determined from galaxy UV luminosity function measurements
  at $z>6$ \citep{2013ApJ...768...71R}.  The escape fraction of
  ionizing radiation increases with redshift.  The clumping factor of
  the high redshift IGM is determined from simulations that are
  similar to our fiducial simulation but with smaller box sizes so
  that the resolution is higher by a factor of 4.  In this model,
  quasars and Population~III stars make a negligible contribution to
  the reionization photon budget.

  The HM12 model agrees reasonably well with the background
  photoionization rate determined from the Ly$\alpha$ forest at $z<6$
  \citep{2009ApJ...703.1416F, 2013MNRAS.436.1023B} and from quasar
  proximity zones at $z\sim 6$ \citep{2011MNRAS.412.1926W,
    2011MNRAS.412.2543C}, albeit with notable differences
  \citep{2015MNRAS.450.4081P, 2015MNRAS.453.2943C}.  The value of the
  electron scattering optical depth to the last scattering surface in
  this model is $\tau=0.084$, which agrees with the WMAP nine-year
  measurements ($\tau=0.089\pm 0.014$; \citealt{2013ApJS..208...19H}).
  This value, however, is inconsistent at more than 1-$\sigma$ level
  with the much lower value of the optical depth reported by Planck
  ($\tau=0.058\pm 0.012$; \citealt{2016arXiv160503507P}).

\item {\bf Late/Default:} CPHB15 found that the rapid disappearance of
  \lya emitters with increasing redshift at $z>6$ suggest a somewhat
  later reionization than predicted by the HM12 model. We have
  therefore chosen the ``Late'' model of CPHB15 as our default
  reionization model, which we call ``Late/Default''.  In this model,
  reionization is complete ($Q_V \sim 1$) at $z = 6$.  The electron
  scattering optical depth in this model is $\tau=0.068$.  This
  reionization history is consistent with constraints derived by
  \citet{2015MNRAS.454L..76M}, and \citet{2016MNRAS.455.4295G}.  As we
  will see the evolution of the neutral hydrogen fraction is also very
  similar with the default model in the suite of radiative transfer
  simulations performed by \citet{2015MNRAS.453.2943C} that fits the
  \lya forest absorption data very well.  \citet{2015MNRAS.453.2943C}
  use hydrodynamical simulation boxes with sufficient resolution to
  resolve \lya forest absorption features and post-process them with
  the cosmological radiative transfer code \textsc{aton}
  \citep{2008MNRAS.387..295A}.  These simulations are able to capture
  the sudden increase of the ionizing photon mean free path and the
  mean photoionization rate due to overlap of H~\textsc{ii} regions
  towards the end of reionization.  As shown by
  \citet{2015MNRAS.453.2943C}, this reionization model also agrees
  well with the photoionization rate measurements at $z<6$
  \citep{2009ApJ...703.1416F, 2011MNRAS.412.1926W,
    2011MNRAS.412.2543C, 2013MNRAS.436.1023B}.  In their radiative
  transfer simulation, the abundance of sources is thereby consistent
  with UV luminosity function measurements at $z>6$
  \citep{2013ApJ...768...71R}.

\item {\bf Very Late:} Finally, we also consider the ``Very Late''
  model introduced by CPHB15.  In this model, reionization completes
  at $z = 6$ (similar to the Late/Default model), but the ionized
  fraction $Q_V$ evolves more rapidly at $z > 6$. The electron
  scattering optical depth in this case is reduced to $\tau = 0.055$.
  This reionization history is consistent with constraints derived by
  \citet{2015MNRAS.454L..76M} and \citet{2016MNRAS.455.4295G} and
  models developed by \citet{2016MNRAS.457.4051K}.
\end{itemize}

Table~\ref{tab:models} summarizes these three models, together with
three variations on the Late/Default model that we consider later in
this paper.  Each of the above three reionization models is specified
by the redshift evolution of the volume-weighted ionization fraction
$Q_V$.  Our simulated ionization field is calibrated to the given
reionization model in two steps.  In the first step, the effective
ionization parameter $\zeta_\mathrm{eff}$ is tuned to get the
volume-weighted ionization fraction predicted by the reionization
model at the corresponding redshift.  In the second step, we obtain
the photoionization rate distribution within the ionized regions by
solving the globally averaged radiative transfer equation
\begin{equation}
  \frac{dQ_V}{dt}=\frac{\dot n_\mathrm{ion}}{n_H}-\frac{Q_V}{t_\mathrm{rec}}
  \label{eqn:reion}
\end{equation}
for the photoionization rate $\Gamma_\mathrm{HI}$.

\begin{figure*}
  \begin{center}
    \includegraphics[width=\textwidth]{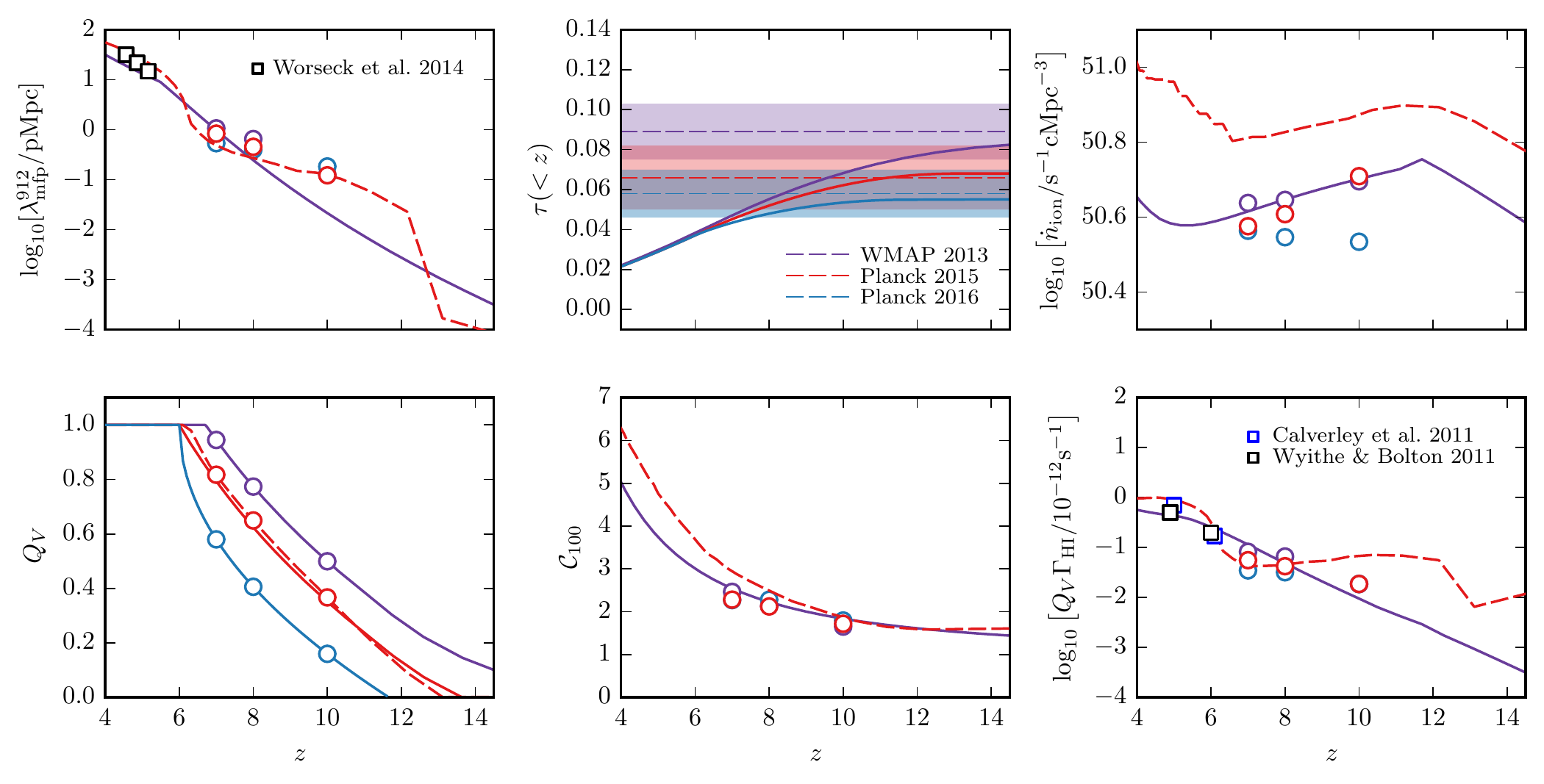}
  \end{center}
  \caption{Calibration of our simulation to various reionization
    models.  The red dashed curves show various quantities from
    \citet{2015MNRAS.453.2943C} which is similar to our Late/Default
    reionization model (shown by the red solid curves).  Purple curves
    show results from \citet{2012ApJ...746..125H} and describe our
    HM12 reionization model.  Blue curves show our Very Late
    reionization model.  Purple, red, and blue circles show values of
    various quantities from our simulation when it is calibrated to
    the HM12, Late/Default, and Very Late reionization models,
    respectively.  Quantities shown are (clockwise from top left): the
    mean free path of ionizing photons, Thomson scattering optical
    depth, rate of ionizing photons, the volume-averaged hydrogen
    photoionization rate, clumping factor of ionized gas with an
    overdensity of less than 100, and the volume-weighted ionization
    fraction.  In the top left panel, black squares show measurements
    of the mean free path of hydrogen-ionizing photons by
    \citet{2014MNRAS.445.1745W}.  In the top middle panel, the dashed
    purple line and the associated purple shaded region show the
    measured value of the Thomson scattering optical depth from WMAP
    \citep{2013ApJS..208...19H} and the associated 1$\sigma$
    uncertainty, respectively.  The red dashed line and shaded region
    show the measured value of the Thomson scattering optical depth
    from the 2015 Planck analysis \citep{2015arXiv150201589P} with its
    associated 1$\sigma$ uncertainty. This value was obtained using
    the Planck CMB power spectra in combination with CMB lensing
    reconstruction.  The blue dashed line and shaded regions show the
    measured value of the Thomson scattering optical depth from the
    2016 Planck analysis \citep{2016arXiv160503507P} with its
    associated 1$\sigma$ uncertainty.  This value was obtained using
    the Planck CMB polarization and temperature data.  In the bottom
    right panel, black and blue squares show measurements of the
    hydrogen photoionization rate by \citet{2011MNRAS.412.1926W} and
    \citet{2011MNRAS.412.2543C}, respectively, from quasar proximity
    zones.}
  \label{fig:review_chardin}
\end{figure*}

Note that in Equation~(\ref{eqn:reion}) the first term on the right
hand side is determined by the average comoving photon emissivity
$\dot n_\mathrm{ion}$ which is related to the photoionization rate by
\citep{2012MNRAS.423..862K, 2013MNRAS.436.1023B}
\begin{equation}
  \dot n_\mathrm{ion} = \frac{\Gamma_\mathrm{HI}Q_V}{(1+z)^2\sigma_H\lambda_\mathrm{mfp}}\left(\frac{\alpha_b + 3}{\alpha_s}\right),
  \label{eqn:term1}
\end{equation}
where $\alpha_s$ is the spectral index of the ionizing sources at
$\lambda<912$~{\AA} and $\alpha_b$ is the spectral index of the
ionizing ``background'' within ionized regions.  The mean free path
$\lambda_\mathrm{mfp}$ also depends on the photoionization rate
$\Gamma_\mathrm{HI}$.  (We will discuss below how the mean free path
is determined.)  The $\Gamma_\mathrm{HI}$ defined above is the
photoionization rate within ionized regions.  The corresponding
globally averaged value is given by $\Gamma_\mathrm{HI} Q_V$.  We use
the same value for the quantity $(\alpha_b+3)/\alpha_s$ as that used
by \citet{2012ApJ...746..125H}.  It is estimated from the model of
\citet{2012ApJ...746..125H} by computing the ratio $\dot
n_\mathrm{ion}\lambda_\mathrm{mfp}/(\Gamma_\mathrm{HI}Q_V)$.

The second term on the right hand side of Equation~(\ref{eqn:reion})
is also related to the photoionization rate.  The recombination time
is given by
\begin{equation}
  t_\mathrm{rec} = \frac{1}{\mathcal{C}\alpha_R\chi\bar n_\mathrm{H}(1+z)^3},
  \label{eqn:term2}
\end{equation}
where
$\mathcal{C}=\langle\rho_\mathrm{HII}^2\rangle/\langle\rho_\mathrm{HII}\rangle^2$
is the clumping factor in the ionized regions, $\alpha_R$ is the
recombination rate, and $\chi=1.08$ is the number of electrons per
hydrogen nucleus (assuming that He~\textsc{i} is completely ionized in
H~\textsc{ii} regions).  The time scale $t_\mathrm{rec}$ is dependent
on the photoionization rate via the clumping factor.

It is possible to use Equations~(\ref{eqn:term1}) and
(\ref{eqn:term2}) to solve Equation~(\ref{eqn:reion}) iteratively for
the photoionization rate at each point in our simulation box if we are
able to calculate the ionized fraction given a photoionization rate
and estimate the mean free path. The resultant combination of gas
density, source distribution, photoionization rate, and the ionization
field are now consistent with the assumed average reionization
history. This procedure breaks down in the post-reionization era when
$Q_V = 1$ and $d Q_V /dt = 0$.  In that case, we assume a value of
$\Gamma_{\rm HI}$ that is consistent with observations at $z\sim
5$--$6$.

\begin{figure*}
    \begin{center}
        \includegraphics*[width=\textwidth]{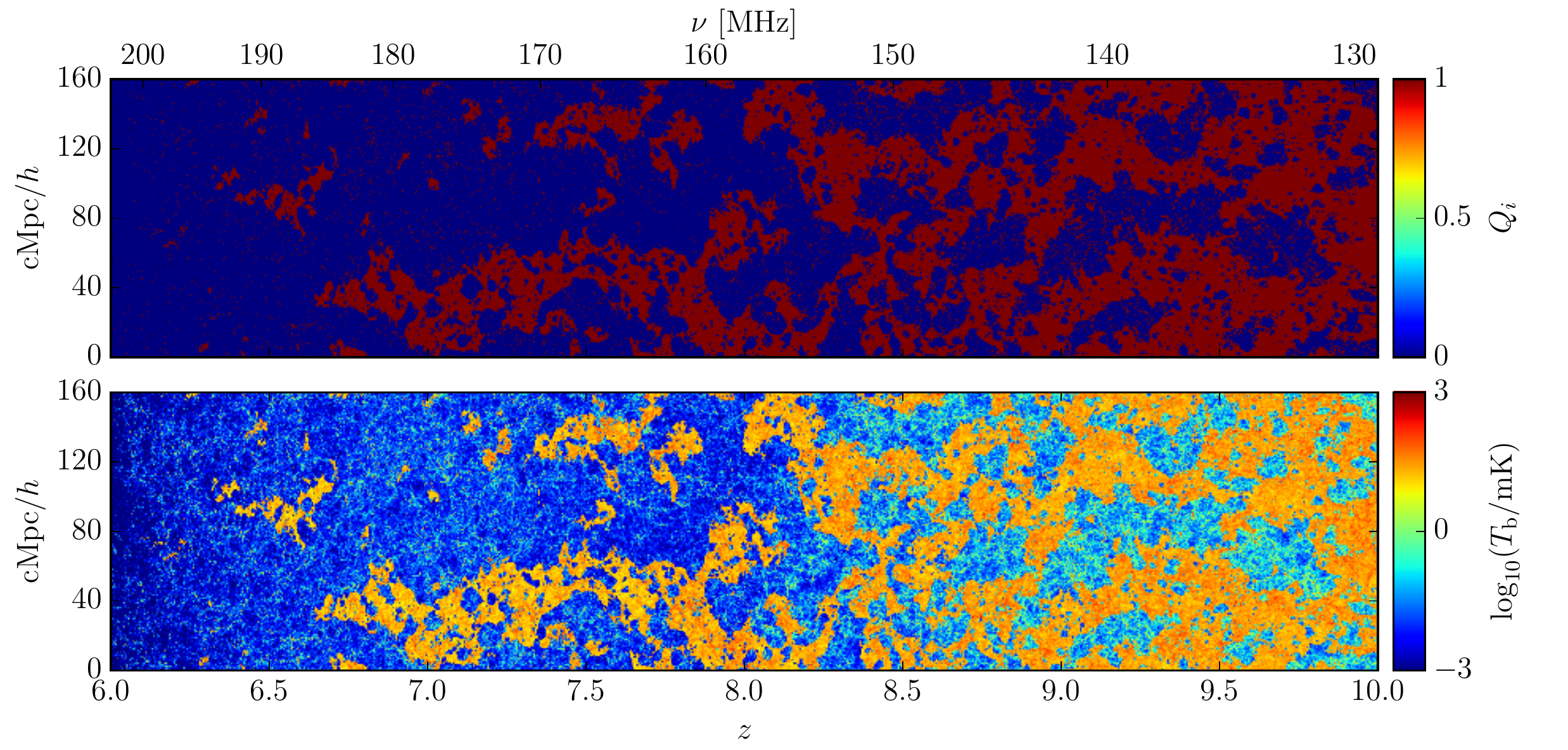}
    \end{center}
    \caption{Light cones of the ionized fraction (top panel) and the
      21~cm brightness temperature (bottom panel) in the Late/Default
      model from $z=6$ to $10$.  The ionized fraction is obtained
      using the excursion set method and our hydrodynamical
      simulations.  21~cm brightness temperature is derived by
      calibrating the ionization field to the Late/Default
      reionization history with assumptions for self-shielding in high
      density regions as suggested by \citet{2013MNRAS.430.2427R}.}
    \label{fig:slices_example}
\end{figure*}

\subsection{Self-shielding}
\label{sec:ss}

Because of the high resolution of our simulation, it is possible to
account for self-shielding and an inhomogeneous photoionization rate
distribution in ionized regions.  Self shielding is implemented in our
simulation during the process of solving Equation~(\ref{eqn:reion})
for the photoionization rate.  For a given photoionization rate
$\Gamma_\mathrm{HI}$ in a grid cell, the neutral fraction
$x_\mathrm{HI}\equiv n_\mathrm{HI}/n_\mathrm{H}$ of the cell is given
by
\begin{equation}
  x_\mathrm{HI}\Gamma^\mathrm{local}_\mathrm{HI}=\chi n_H(1-x_\mathrm{HI})^2\alpha_\mathrm{R},
  \label{eqn:ioneq}
\end{equation}
where $\Gamma^\mathrm{local}_\mathrm{HI}$ is the local photoionization
rate in the cell.  This density-dependent photoionization rate is
obtained from the background photoionization rate $\Gamma_\mathrm{HI}$
using the fitting function obtained by \citet{2013MNRAS.430.2427R}
from radiative transfer simulations
\begin{equation}
  \frac{\Gamma^\mathrm{local}_\mathrm{HI}}{\Gamma_\mathrm{HI}}=0.98\left[1+\left(\frac{\Delta_\mathrm{H}}{\Delta_\mathrm{ss}}\right)\right]^{-2.28}+0.02\left[1+\frac{\Delta_\mathrm{H}}{\Delta_\mathrm{ss}}\right]^{-0.84}.
  \label{eqn:gammass}
\end{equation}
Here $\Delta_\mathrm{ss}$ is a self-shielding density threshold given
by \citep{2001ApJ...562L..95S, 2005ApJ...622....7F,
  2013MNRAS.430.2427R}
\begin{multline}
  \Delta_\mathrm{ss} = 36\left(\frac{\Gamma_\mathrm{HI}}{10^{-12}~\mathrm{s}^{-1}}\right)^{2/3}\left(\frac{T}{10^4~\mathrm{K}}\right)^{2/15}\\\times\left(\frac{\mu}{0.61}\right)^{1/3}\left(\frac{f_e}{1.08}\right)^{-2/3}\left(\frac{1+z}{8}\right)^{-3},
  \label{eqn:deltass}
\end{multline}
where $T$ is the gas temperature, $\mu$ is the mean molecular weight,
and $f_e=n_e/n_\mathrm{H}$ is the ratio of free electron and hydrogen
number densities.  We assume $T=10^4$ K in ionized regions.

Figure~\ref{fig:skewers} illustrates the effect of self-shielding by
showing the distribution of neutral hydrogen density, total gas
density, and the ionization field along a line of sight through the
simulation box at $z=7$ for the Late/Default model.  Panel (a) shows
the H~\textsc{i} density distribution in the absence of
self-shielding.  Panel (b) shows the H~\textsc{i} density distribution
with self-shielding.  Panels (c) and (d) show the total gas
overdensity and the large scale ionization field along this line of
sight.  In the absence of self-shielding the photoionization rate is
constant across the large ionized region from $x\sim 25$ to $150$
$h^{-1}$cMpc and the neutral hydrogen fraction is significantly
non-zero only outside this region.  In contrast, panel (b) shows the
neutral hydrogen distribution when self-shielding is applied following
the prescription in Equations~(\ref{eqn:ioneq}) and
(\ref{eqn:gammass}).  Locations within ionized regions now show high
neutral hydrogen fraction if they have sufficiently high density as
seen, e.g., at around $x\sim 100$ and 120 $h^{-1}$cMpc.  Our
calibration technique allows us to account for these self-shielded
regions.

With a chosen self-shielding criterion, we obtain a neutral hydrogen
distribution across the box.  This can then also be used to calculate
the mean free path of ionizing photons.  To estimate  the mean free path, we
calculate the mean transmission at 912 {\AA} across a large number of
sightlines through the box.  Figure~\ref{fig:flux} shows the mean
transmission obtained from 1000 sightlines in the box at $z=7$
corresponding to the two self-shielding cases shown in
Figure~\ref{fig:skewers}.  The mean free path $\lambda_\mathrm{mfp}$
is then just obtained by fitting the mean transmission by
\begin{equation}
  \langle \exp\left(-\tau_{912}\right)\rangle = F_0\exp\left(-\frac{x}{\lambda_\mathrm{mfp}}\right),
  \label{eqn:mfp}
\end{equation}
where $x$ is the position along a sightline, as expected for radiative
transfer in a highly ionized medium \citep{1985rpa..book.....R}.  

Figure~\ref{fig:review_chardin} shows the result of calibrating our
simulation to the HM12 (purple curves and symbols), Late/Default (red
curves and symbols), and Very Late (blue curves and symbols)
reionization models at redshifts $z=7, 8,$ and $10$.  The
volume-weighted ionization fraction, $Q_V$, in the simulation is
matched to the model by tuning the effective ionization emissivity
parameter $\zeta_\mathrm{eff}$.  We then get a good agreement between
the mean free path and clumping factor in the simulation and the
models.  This is reflected in the good agreement on the total ionizing
photon emission rate, $\dot n_\mathrm{ion}$ and the photoionization
rate.  Note that Figure~\ref{fig:review_chardin} shows the clumping
factor of regions with overdensity less than 100 for consistency.  In
our analysis, we use the clumping factor calculated throughout the
ionized regions.  At redshift $10$, the low value of the
photoionization rate in Equation~(\ref{eqn:deltass}) results in an
unrealistically low value of $\Delta_\mathrm{ss}$, which can
potentially self-shield all the gas in the ionized regions.  This
problem has been noted by CPHB15. Here we restrict
$\Delta_\mathrm{ss}\geq 10$ while calibrating our simulation at this
redshift.  As already discussed the reionization history in our
Late/Default model is very similar to the default model in
\citet{2015MNRAS.453.2943C} . Note that the required ionizing
emissivity (upper right panel) in the simulations of
\citet{2015MNRAS.453.2943C} depends on resolution due to
recombinations in the host haloes of the ionizing sources.  The
simulations of \citet{2015MNRAS.453.2943C} are furthermore
monochromatic.  The ionizing emissivities therefore differ between
their work and our model, even though the evolution of the ionized
volume fraction is very similar.

\section{Cosmological 21~cm Signal}
\label{sec:21cm}

\begin{figure*}
  \begin{center}
    \includegraphics*[width=\textwidth]{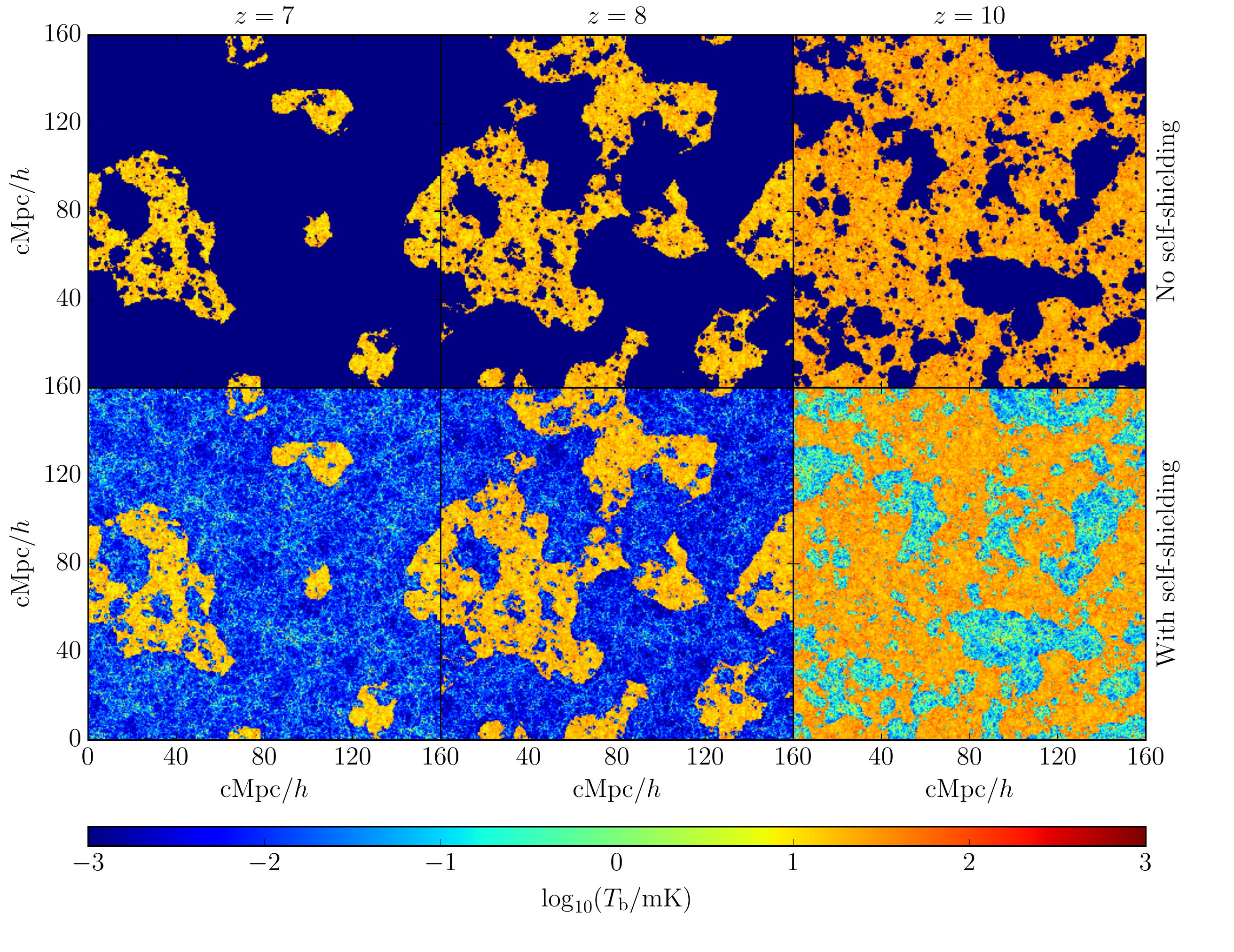}
  \end{center}
  \caption{Slices of the 21~cm brightness temperature distribution at
    $z=7, 8,$ and $10$ (left, middle, right columns, respectively) in
    the Late/Default model, with (bottom row) and without (top row)
    self-shielding.  Self-shielding adds structure in ionized
    regions.}
  \label{fig:tbslices}
\end{figure*}

\begin{figure*}
    \begin{center}
      \includegraphics*[width=\textwidth]{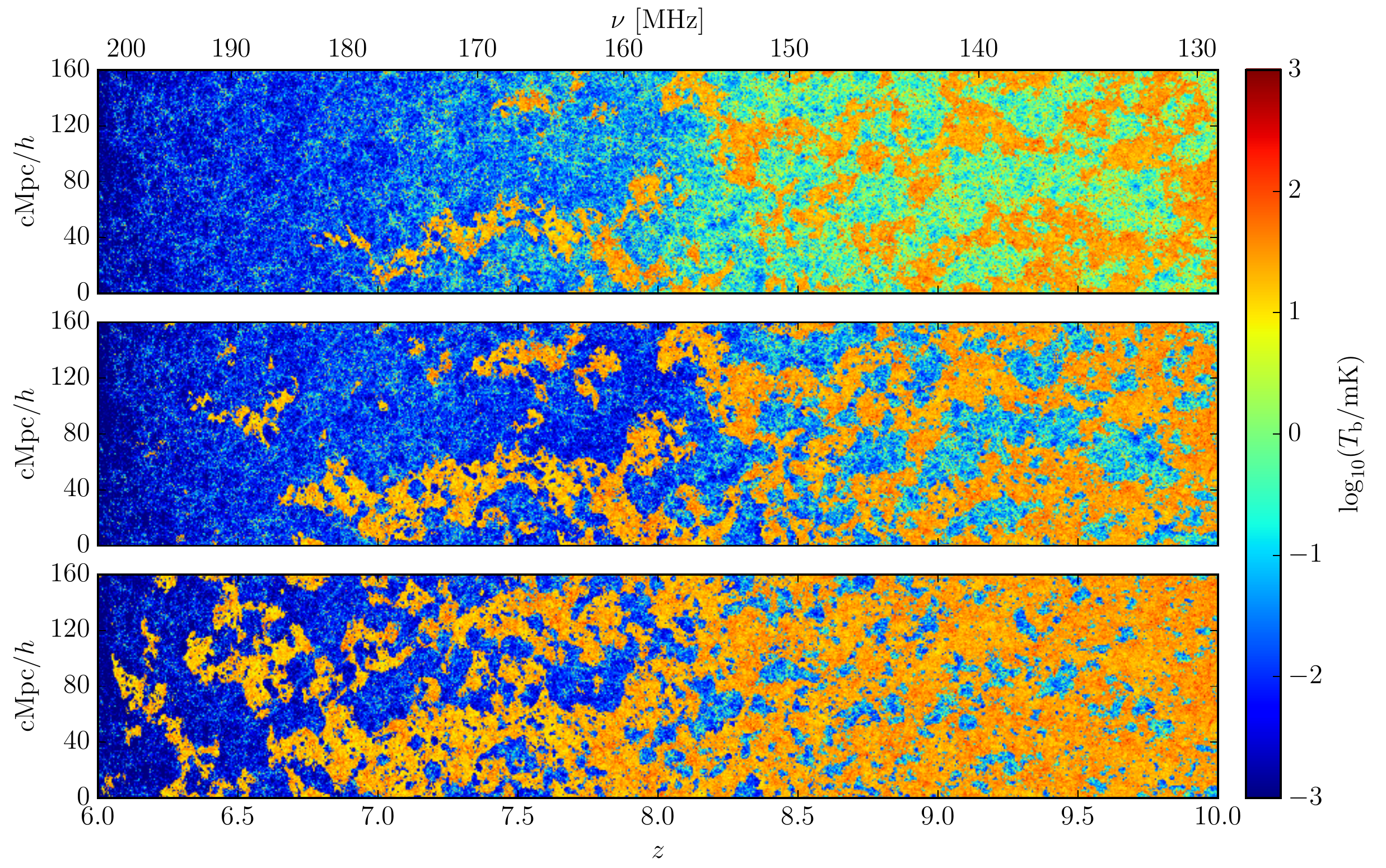}
    \end{center}
    \caption{The 21~cm brightness distribution in the HM12 (top
      panel), Late/Default (middle panel), and Very Late (bottom
      panel) models.}
    \label{fig:lightcones}
\end{figure*}

The calibrated ionization field calculated above can now be used to
derive the 21~cm brightness distribution from the epoch of
reionization.  Due to the calibration procedure described above, this
distribution accounts for inhomogeneous
recombinations in self-shielded regions of the IGM and is consistent
with a variety of other constraints on reionization.

\subsection{21~cm brightness temperature}

The 21~cm brightness temperature can be approximated as
\begin{equation}
  T_b(\mathbf{x})=\overline T_b x_\mathrm{HI}(\mathbf{x})\Delta(\mathbf{x}),
  \label{eqn:tb}
\end{equation}
where the mean temperature $\overline T_b\approx 22 \mathrm{mK}
[(1+z)/7]^{1/2}$ \citep{2009MNRAS.394..960C}.  The above relation does
not account for the fluctuations in the spin temperature, i.e., it
implicitly assumes that the spin temperature is much greater than the
CMB temperature and that the Ly$\alpha$ coupling is sufficiently
complete throughout the IGM.  These conditions are likely met in the
redshift range considered here, when the global ionized fraction is
greater than a few per cent \citep{2012RPPh...75h6901P,
  2015MNRAS.447.1806G}.

Figure~\ref{fig:slices_example} shows light cones through the
ionization field and 21~cm brightness distribution in our simulation
from $z=6$ to $10$ for the Late/Default reionization model.  The
horizontal span of the figure is $\sim 1200$ $h^{-1}$cMpc,
corresponding to the comoving distance from $z=6$ to $10$.  As
expected, the neutral regions are the brightest in 21~cm, but there
are also self-shielded regions with $T_b\sim 1$--$10$ mK within the
ionized regions.  The number of 21~cm bright regions increases with
redshift.  It should be noted here that given a reionization history
the resultant morphology of the ionized regions in this model is not
unique; it is dictated by our assumptions that the total ionizing
photon contribution of a halo is proportional to its mass, our
assumptions about the spectral index of each source, and the mass
range of haloes considered.  We consider these issues in greater detail
in Section~\ref{sec:ss_effect} below.

To understand the influence of recombinations on 21~cm brightness,
Figure \ref{fig:tbslices} shows the distribution of the 21~cm
brightness temperature in our Late/Default model at redshifts $z=7,
8,$ and $10$ with and without self-shielding.  To obtain the
brightness distribution without self-shielding, we repeat the
calibration procedure described in Section~\ref{sec:calibration} but
force the photoionization rate to be uniform within the ionized
regions.  For a given volume weighted ionization fraction, this
typically results in a higher photoionization rate than the
self-shielded case, as the mean free path of ionizing photons is
significantly larger in the absence of self-shielded sinks
(Figure~\ref{fig:flux}).  Note that because our calibration procedure
fixes the volume weighted ionization fraction by construction, it
leaves the size of the ionized regions unchanged when self-shielding
is introduced.  As a result, panels in the top and bottom rows of
Figure \ref{fig:tbslices} have very similar large scale structure.
However, the brightness temperature within ionized regions is
significantly higher in the simulations with self-shielding due to the
presence of neutral hydrogen in dense regions.  As gas density
increases as $(1+z)^3$ and the photoionization rate decreases, this
effect is relatively stronger at higher redshifts.

The light cones of the 21~cm brightness distribution for the three
reionization histories with self-shielding are shown in
Figure~\ref{fig:lightcones}.  The evolution of ionized regions in the
three models is as expected, with an early reionization in the HM12
and rapid but late reionization in the Very Late model.  The ionized
regions in the HM12 model are relatively brighter than those in the
Late/Default and Very Late models because the photoionization rate in
the HM12 models falls much more rapidly at high redshifts.  The effect
of self-shielding is visible in the ionized regions in all three
models.

\subsection{Power spectra}

\begin{figure*}
  \begin{center}
    \includegraphics[width=\textwidth]{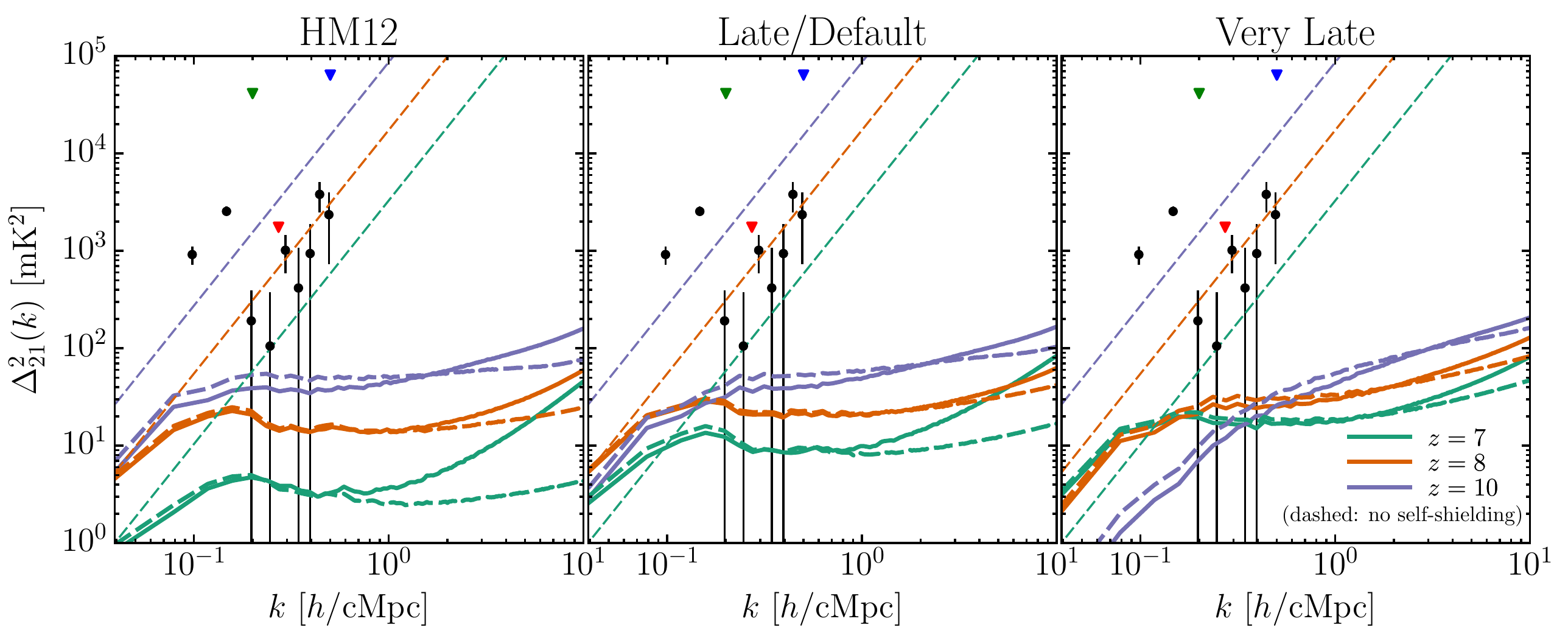}
  \end{center}
  \caption{21~cm power spectra at $z=7$ (green curves), $8$ (orange
    curves) and $10$ (purple curves) in the three reionization models:
    HM12 (left panel), Late/Default (middle panel), and Very Late
    (right panel).  Solid curves show the power spectra when
    self-shielding in ionized regions is accounted for; dashed curves
    show the power spectra without self-shielding.  Blue triangles
    show upper limits on the power spectrum at $z=8.6$ from GMRT
    \citep{2013MNRAS.433..639P}.  Green triangles show upper limits at
    $z=9.5$ from MWA \citep{2014PhRvD..89b3002D}.  Red triangles show
    upper limits at $z=7.7$ from the 32-element deployment of PAPER
    \citep{2014ApJ...788..106P}.  Black points with error bars show
    measurements at $z=8.4$ by the 64-element deployment of PAPER
    \citep{2015ApJ...809...61A} with their 2-$\sigma$ uncertainties.
    Dashed lines show our estimated sensitivity limits for the
    64-element deployment of PAPER for same integration time as
    \citet{2015ApJ...809...61A} at $z=7$ (green), $8$ (orange) and
    $10$ (purple).}
  \label{fig:ps_evolution}
\end{figure*}

Spatial fluctuations in the 21~cm brightness distribution are
conveniently characterized by their power spectrum
\citep{2006PhR...433..181F}.  Figure~\ref{fig:ps_evolution} shows the
spherically averaged real-space 21~cm power spectra at $z=7, 8$ and
$10$ in our simulation for the HM12, Late/Default, and Very Late
reionization histories.  We define the power spectrum as
\begin{equation}
  \Delta_{21}^2(k) = \frac{k^3\langle\tilde{T_b^2}(k)\rangle}{2\pi^2},
\end{equation}
where $\tilde{T_b}(k)$ is the Fourier transform of the brightness
temperature defined in Equation~\ref{eqn:tb} and the average is over
the simulation box.  Also shown in Figure~\ref{fig:ps_evolution} are
the measurements and upper limits from various experiments such as the
Giant Metrewave Radio Telescope (GMRT, $z=8.6$;
\citealt{2013MNRAS.433..639P}), Murchison Widefield Array (MWA,
$z=9.5$; \citealt{2014PhRvD..89b3002D}), the 32-element deployment of
the Precision Array for Probing the Epoch of Reionization (PAPER,
$z=7.7$; \citealt{2014ApJ...788..106P}), and the 64-element deployment
of PAPER ($z=8.4$; \citealt{2015ApJ...809...61A}).  The dashed lines
show our estimated sensitivity limits for the 64-element deployment of
PAPER for the same integration time as \citet{2015ApJ...809...61A} at
$z=7$, $8$ and $10$.  We describe how these are derived in
Section~\ref{sec:obs} below.

The evolution of the power spectrum is qualitatively familiar
\citep{2006PhR...433..181F, 2012RPPh...75h6901P}.  At large scales the
power spectrum amplitude is around 1 to 20 mK$^2$, depending on the
model.  The power spectrum amplitude continuously increases with
redshift in the HM12 model \citep{2008ApJ...680..962L,
  2009MNRAS.394..960C, 2016MNRAS.457.1550H} due to increasing presence
of dense, neutral regions.  In the Late/Default model, the large scale
power first increases from $z=7$ to $8$ and then decreases at $z=10$.
This is the ``rise and fall'' signature of reionization that is
visible in the 21~cm power spectrum in the Late/Default model because
of the relatively late reionization.  This effect is further enhanced
in the Very Late model, in which the power spectrum drops from $z=7$
to $10$. At small scales the power spectrum follows the matter
spectrum while at intermediate to large scales it deviates from the
matter spectrum due to the clustering of ionized regions
\citep{2006PhR...433..181F}.  Our predicted 21~cm power spectra are
about a factor of three lower than the lowest current experimental
upper limits.

Figure~\ref{fig:ps_evolution} also shows the effect of self-shielded
regions on the 21~cm power spectra in the three models.  Dashed curves
in this figure show the 21~cm power when no self-shielding is assumed.
In general the effect of self-shielding is to increase 21~cm power at
small scales.  This is due to increased fluctuations in the 21~cm
brightness resulting from the presence of neutral hydrogen in dense
regions as seen in Figure~\ref{fig:tbslices}.  This increase in power
is seen at all three redshifts in Figure~\ref{fig:ps_evolution} at
scales corresponding to $k\gtrsim 1$ cMpc$^{-1}h$.  The enhancement of
power generally decreases with increasing redshift between $z=7$ and
$10$.  At high redshift smaller volume filling factors of ionized
regions reduces the impact of self-shielding.  In the Late/Default
model, at $k=10$ cMpc$^{-1}h$ the power increment is by about a factor
of $\sim 2$ at $z=10$ but it rises to factor of $\sim 5$ at $z=7$.

The effect of self-shielding at large scales is opposite to that at
small scales.  Here, power is reduced as a result of the reduction in
the contrast between the brightness of ionized and neutral regions.
This is in qualitative agreement with previous findings in the
literature \citep{2014MNRAS.440.1662S, 2015arXiv151008767K,
  2016MNRAS.457.1550H}.  Also, contrary to small scales, the effect of
self-shielding on large scales generally increases with increasing
redshift.  In the Late/Default model, at $k=0.1$ cMpc$^{-1}h$, the
effect at $z=7$ is about 10\%, while at $z=10$ the power is reduced by
a factor of 30\%.  Thus, predictions of standard excursion set models
of the 21~cm brightness distribution at large scales during the epoch
of reionization grow successively worse at higher redshifts.

\begin{table*}
  \begin{center}
    \begin{tabular}{lccccc}
      \hline
      Parameter & PAPER & MWA & LOFAR & HERA & SKA1-LOW \\
      \hline
      Number of antennae ($N_\mathrm{ant}$) & 132 & 126 & 48 & 547 & 512 \\
      Effective collecting area ($A_\mathrm{eff}/m^2$) & 4.0 & 13.8 & 526.0 & 154.0 & 962.0 \\
      Maximum baseline ($b_\mathrm{max}/m$) & 192.3 & 2401.9 & 3475.6 & 400.0 & 40286.8 \\
      Minimum baseline ($b_\mathrm{min}/m$) & 4.0 & 3.72 & 22.92 & 14.0 & 16.8 \\
      \\
      Largest observable scale at $z=7$ ($k_\perp^\mathrm{min}/10^{-4}\mathrm{cMpc}^{-1}h$) & 3.8 & 3.5 & 21.8 & 13.3 & 16.0 \\ 
      Largest observable scale at $z=8$ ($k_\perp^\mathrm{min}/10^{-4}\mathrm{cMpc}^{-1}h$) & 3.3 & 3.1 & 19.0 & 11.6 & 13.9 \\ 
      Largest observable scale at $z=10$ ($k_\perp^\mathrm{min}/10^{-4}\mathrm{cMpc}^{-1}h$) & 2.6 & 2.4 & 14.9 & 9.1 & 10.9 \\
      \hline
    \end{tabular}
  \end{center}
  \caption{Parameters used for the five experiments considered in this
    paper.  Note that number of antennas in HERA in the final
    instrument design is 350 \citep{2016arXiv160607473D}.  We use SKA
    parameters obtained by \citet{2016MNRAS.460..827G} which broadly
    agrees with the baseline distribution given in the latest SKA1-LOW
    configuration document
    (http://astronomers.skatelescope.org/documents/; Document number
    SKA-SCI-LOW-001; date 2015-10-28).}
  \label{tab:experiments}
\end{table*}

The effect of self-shielding is similar in the HM12 and Very Late
models: power is enhanced at small scales and reduced at large scales.
The higher the volume filling fraction of ionized regions, the larger
is the effect on small scales and smaller is the effect at large
scales.  At $z=7$ the volume-weighted ionization fraction, $Q_V$, in
the HM12 model is 0.94, while it is 0.82 in the Late/Default model and
0.58 in the Very Late model.  The corresponding difference in the
volume filling fraction of neutral hydrogen in the Late/Default and
Very Late models increases the predicted 21~cm brightness temperature
in the Very Late model as compared to the Late/Default model and
decreases it in the HM12 model.  This is in turn reflected in the
amplitude of the 21~cm power spectra for these models at $z=7$ as seen
in Figure~\ref{fig:ps_evolution}.  At $z=7$ the shape of the power
spectrum curves for these three reionization models is very similar
but the amplitude increases in the Very Late model relative to the
Late/Default model.  The power spectrum amplitude in the HM12 is lower
than that in the Late/Default model.  The effect of self-shielding is
also qualitatively similar to that in the Late/Default case: there is
a $\sim 10\%$ decrease in the power at large scales ($k\sim 0.1$
cMpc$^{-1}h$) and a significant increase in the power at small scales
($k\sim 10$ cMpc$^{-1}h$).  The magnitude of this effect is highest in
the HM12 model.  This can again be understood from the difference in
the ionization fractions in the three models.

At redshifts $z=8$ and $10$, the ionization fraction in the Very Late
model is much lower than in the Late/Default model.  As a result, at
$z=8$ the 21~cm power spectrum in the Very Late model already has a
lower amplitude than in the Late/Default model at large scales ($k\sim
0.1$ cMpc$^{-1}h$).  At $z=10$, both Late/Default and Very Late model
predict lower power spectrum amplitudes relative to the HM12 model at
large scales.  The result of this behaviour is that in the Very Late
model the large scale ($k\sim 0.1$ cMpc$^{-1}h$) power continuously
drops $z=7$ to $z=10$ unlike the power spectrum in the Late/Default
model, which shows a rise-and-fall behaviour at large scales.  For the
Very Late model, the fall in 21~cm brightness does not occur until
$z<7$.  In the HM12 model, on the other hand, $Q_V$ is large enough to
push the rise-and-fall signature to redshifts higher than 10.  In this
model the 21~cm power continuously increases from $z=7$ to $z=10$.  A
general result is that there is up to 30\% decrement in large scale
power due to self-shielding.

Note that at post-reionization redshifts ($z<6$), models without
self-shielding predict uniformly zero 21~cm signal.  At these
redshifts, significant amounts of neutral hydrogen are only present in
self-shielded systems.  However, the overall amplitude of the 21~cm
power spectrum is very low at these redshifts and is unlikely to be
detected by any of the five experiments considered here.

\begin{figure*}
  \begin{center}
    \includegraphics[width=\textwidth]{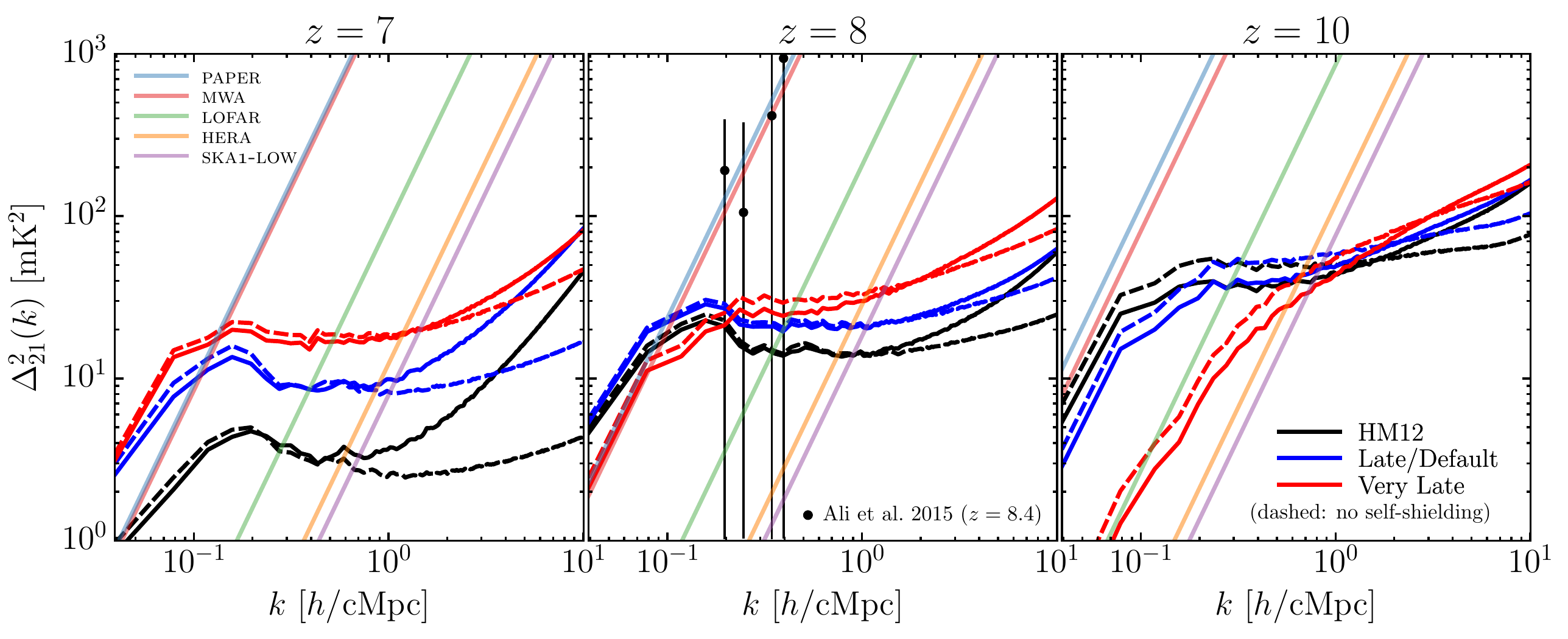}
  \end{center}
  \caption{Effect of self-shielding on power spectra on the 21~cm
    power spectrum at $z=7$ (left panel), $8$ (middle panel), and $10$
    (right panel) in our three models.  Black, blue, and red curves
    show power spectra from the HM12, Late/Default, and Very Late
    models, respectively.  Solid curves show power spectra when
    self-shielding is accounted for.  Dashed curves show power spectra
    without self-shielding.  Black points with error bars show
    measurements at $z=8.4$ by the 64-element deployment of PAPER
    \citep{2015ApJ...809...61A} with their 2-$\sigma$ uncertainties.
    Dashed curves show our estimates of the sensitivities for PAPER
    (blue), MWA (red), LOFAR (green), HERA (orange), and SKA1-LOW
    (purple).}
  \label{fig:ps_3zs}
\end{figure*}

\subsection{Detecting the 21~cm Power Spectrum}
\label{sec:obs}

We study here the observational detectability of the 21~cm power
spectrum, for five experiments: PAPER \citep{2014ApJ...788..106P}, MWA
\citep{2013PASA...30...31B, 2013PASA...30....7T}, Low Frequency Array
(LOFAR; \citealt{2013A&A...556A...2V}; \citealt{2014ApJ...782...66P}),
Hydrogen Epoch of Reionization Array (HERA;
\citealt{2014ApJ...782...66P}, \citealt{2016arXiv160607473D}), and the
low frequency instrument from Phase 1 of the Square Kilometre Array
(SKA1-LOW; http://astronomers.skatelescope.org).  These are listed in
Table~\ref{tab:experiments}.  We assume a system temperature of
\citep{2007isra.book.....T}
\begin{equation}
  T_\mathrm{sys}=60~\mathrm{K}\left(\frac{300~\mathrm{MHz}}{\nu_c}\right)^{2.25},
\end{equation}
and calculate the thermal noise power for an integration over
180~days, assuming a bandwidth of 6~MHz, an observing time of 6~hr per
day, and a mid-latitude location.

We follow the method described by \citet{2012ApJ...753...81P} to
calculate experimental sensitivities, which we briefly summarize
below.  The $uv$ coverage of an interferometric array, obtained by
accounting for Earth rotation synthesis of the array's baselines, is
binned in $uv$ pixels.  Each pixel then corresponds to an independent
sampling of a transverse $k_\perp$ mode of the cosmological 21~cm
brightness distribution.  For each such sampling, the array measures a
range of line-of-sight $k_\parallel$ modes depending on the bandwidth
and frequency resolution.  The thermal white noise power for each mode
is then given by \citep{2006ApJ...653..815M, 2012ApJ...753...81P}
\begin{equation}
  \Delta_\mathrm{thermal}^2(k)\approx X^2Y\frac{k^3}{2\pi^2}\frac{\Omega}{2t}T^2_\mathrm{sys},
  \label{eqn:powerpermode}
\end{equation}
where $k = (k_\perp^2 + k_\parallel^2)^{1/2}$, $\Omega$ is the field
of view of an element of the array, and $t$ is the total integration
time for this $k$-mode.  The field of view is given by
$\lambda^2/A_\mathrm{eff}$ where $\lambda=21~\mathrm{cm}(1+z)$ and
$A_\mathrm{eff}$ is the effective area of an interferometric element.

The cosmological quantities $X$ and $Y$
convert from angles and frequencies to comoving distance,
respectively, and are given by \citep{2012ApJ...753...81P}
\begin{equation}
  X\approx 1.9 \frac{h^{-1}\mathrm{cMpc}}{\mathrm{arcmin}} \left(\frac{1+z}{10}\right)^{0.2},
\end{equation}
and
\begin{equation}
  Y\approx 11.5 \frac{h^{-1}\mathrm{cMpc}}{\mathrm{MHz}} \left(\frac{1+z}{10}\right)^{0.5}\left(\frac{\Omega_mh^2}{0.15}\right)^{-0.5}.
\end{equation}
The factor of two in Equation~(\ref{eqn:powerpermode}) assumes that
two orthogonal polarizations are measured.  To obtain the total
thermal noise power, the power from individual modes, given by
Equation~(\ref{eqn:powerpermode}), can be added in quadrature
suitably.  We assume that every baseline can contribute to the
measurement of each $k$-mode, i.e., that the range of $k_\parallel$ is
broad enough.  When multiple non-instantaneously redundant
measurements are made, measurements of a $k$-mode can be added up in
quadrature thereby reducing the uncertainty in power by the square
root of number of measurements.  On the other hand, multiple
instantaneously redundant measurements of a $k$-mode are equivalent to
coherent integration of the temperature measurement.  This reduces the
uncertainty in power with increasing number of measurements.  In this
paper, we add all sampling in quadrature.  This is in a sense the
worst case estimate of the thermal noise.

We first combine redundant measurements in quadrature in each $k$-bin
in which the power spectrum is measured.  For logarithmic bins of
width $\Delta\ln k$, this modifies the thermal noise power of
Equation~(\ref{eqn:powerpermode}) so that
\begin{equation}
  \Delta_\mathrm{thermal}^2(k)\approx X^2Y\frac{k^{5/2}}{2\pi^2}\left(\frac{1}{B}\right)^{1/2}\left(\frac{1}{\Delta\ln k}\right)^{1/2}\frac{\Omega}{2t}T^2_\mathrm{sys},
  \label{eqn:powerperbin}
\end{equation}
where $B$ is the bandwidth of the observation, which decides the total
number of $k$-modes observed for a given resolution.  In this paper,
we assume $B=6$ MHz for all experiments. 

A second combination is performed over measurements of the same
$k$-mode by different baselines as they are moved into suitable
$uv$-pixels by Earth's rotation.  In order to do this calculation, in
what follows, we assume that $N_\mathrm{ant}$ antennas are distributed
uniformly up to a maximum baseline $b_\mathrm{max}$.  Thus all
redundant measurements are added in quadrature and instantaneously
redundant measurements are ignored.  Since we assume that each mode
can be observed by every baseline, that introduces a term $\propto
1/\sqrt{N}$, where $N$ is number of baselines.  Following
\citet{2012ApJ...753...81P}, we first sum the sensitivity over rings
of $uv$-pixels and then sum over all such rings.  This results in a
power spectrum given by
\begin{multline}
  \Delta_\mathrm{thermal}^2(k)\approx X^2Y\frac{k^{5/2}}{2\pi^2}\left(\frac{1}{B}\right)^{1/2}\left(\frac{1}{\Delta\ln k}\right)^{1/2}\\\times\frac{\Omega}{2t}T^2_\mathrm{sys}\frac{u_\mathrm{max}^{1/2}}{N}\frac{1}{\Omega^{1/4}}\frac{1}{t_\textrm{per-day}^{1/2}},
  \label{eqn:thermalpower}
\end{multline}
where $u_\mathrm{max}$ is the maximum baseline $b_\mathrm{max}$ in
wavelength units.  We assume $t_\textrm{per-day}=6$~hr for 120 days.
Thus the sensitivity is proportional to $k^{5/2}$.

The thermal noise power spectrum calculated using
Equation~(\ref{eqn:thermalpower}) determines the power spectrum
sensitivity of 21~cm experiments.  Figure~\ref{fig:ps_3zs} shows the
sensitivities of the five experiments described in
Section~\ref{sec:21cm} above.  Current and upcoming 21~cm experiments
are only sensitive to large scales due to limited baselines.  As a
result, accounting for self-shielding predicts a decrease in the
signal to noise ratio for these experiments compared to the
predictions from simulations with a more limited dynamic range.  None
of the experiments are sensitive to 21~cm power for $k\gtrsim 1$
cMpc$^{-1}h$.  SKA1-LOW and HERA have the highest sensitivities
primarily due to large number of antenna elements.  The signal to
noise ratio is about 100 for these two experiments $k\sim 0.1$
cMpc$^{-1}h$.  At $z=7$ and $8$, self-shielding has little effect on
the signal to noise ratio, but it is reduced by about a factor of less
than two at $z=10$ due to the reduction of signal power as discussed
in the previous section.  LOFAR has sensitivity for scales
corresponding to $k\lesssim 0.2$ cMpc$^{-1}h$.  At $k\sim 0.1$
cMpc$^{-1}h$, the signal to noise ratio for LOFAR is $\sim 10$ and the
fractional reduction at $z=10$ is significant.  PAPER and MWA are the
least sensitive of the five experiments that we consider in this
paper, due to relatively small number of antennas and shorter
baselines.

Figure~\ref{fig:ps_evolution} shows our estimated sensitivity limits
for the 64-element deployment of PAPER for the same integration time
as \citet{2015ApJ...809...61A} at $z=7$, $8$ and $10$ together with
their reported measurements.  The largest scale observed by the
experiments considered here corresponds to the smallest baseline in
the array.  This is listed in Table~\ref{tab:experiments} and is of
the order of $10^{-4}\mathrm{cMpc}^{-1}h$ for all experiments.  Note
that the sample variance from the limited number of $k$-modes measured
in the survey volume also limits the sensitivity of the experiment.
The sample variance scales as $\Delta^2(k)/\sqrt{N}$ and, due to the
small amplitude of the power spectrum, is smaller ($\lesssim 1$
mK$^2$) than the thermal noise at all redshifts for all experiments
considered here.  Note that we assume perfect foreground subtraction
in this discussion.  Foreground subtraction will reduce the
experimental sensivity \citep{2009A&A...500..965B,
  2013ApJ...768L..36P, 2014PhRvD..89b3002D}.  Due to the relatively
smooth dependence of astrophysical foregrounds on frequency, this
reduction in sensitivity particularly affects small $k$ values.

\begin{figure*}
  \begin{center}
    \includegraphics*[width=\textwidth]{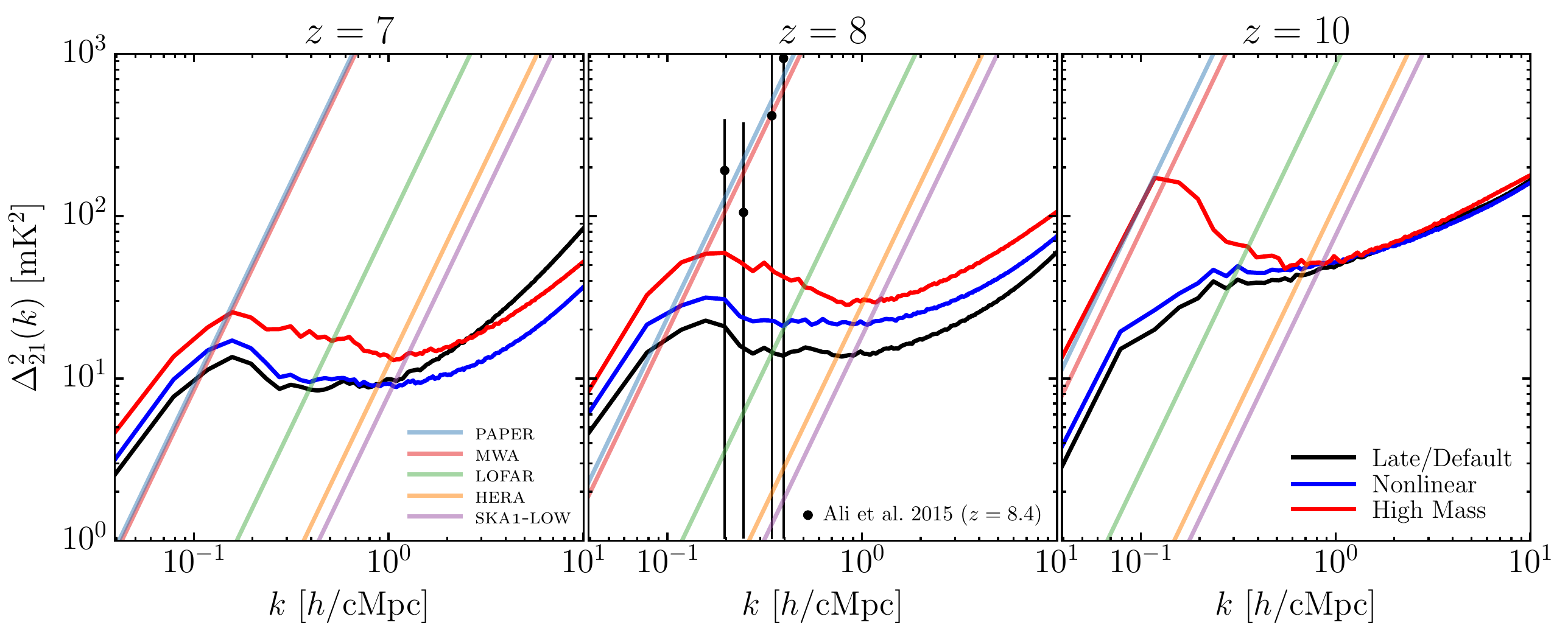}
  \end{center}
  \caption{Power spectra of the 21~cm brightness temperature
    distribution in three models with the Late/Default reionization
    history at $z=7$ (left panel), $8$ (middle panel) and $10$ (right
    panel).  Black curves correspond the Late/Default model, blue
    curves show the Nonlinear model, and red curves show the High Mass
    model.  Experimental sensitivities are also shown: PAPER (blue),
    MWA (red), LOFAR (green), HERA (orange), and SKA1-LOW (purple).
    Black points with error bars show measurements at $z=8.4$ by the
    64-element deployment of PAPER \citep{2015ApJ...809...61A} with
    their 2-$\sigma$ uncertainties.}
\label{fig:ps_nonstd}
\end{figure*}

\section{Model variations}
\label{sec:ss_effect}

So far we have assumed that the total number of ionizing photons
produced by a halo, $N_\gamma$, is proportional to its mass $M$.  We
have also assumed that sources exist in haloes down to a mass of $\sim
10^8$ M$_\odot$.  In this section, we discuss the influence of these
assumptions on our results.

\subsection{Nonlinear dependence of ionization photon emission rate on halo mass}
\label{sec:nonlin}

The assumption that the number of ionizing photons emitted by a halo,
$N_\gamma(M)$, defined in Equation~(\ref{eqn:ngamma}), scales linearly
with halo mass can be a bad approximation if, say, the star formation
rate (SFR) or the escape fraction of ionizing photons from a halo
scales nonlinearly with the halo mass.  Galactic outflows and
photoionization feedback can influence the dependence of star
formation rate on halo mass.  In their cosmological radiation
hydrodynamical simulations, \citet{2011MNRAS.410.1703F} found that
feedback not only suppresses star formation in low mass haloes but
also affects star formation rate in high mass haloes via hierarchical
merging.  They found that the SFR scales as $M_\mathrm{halo}^{1.41}$
for $M_\mathrm{halo}$ between $\sim 10^8$~M$_\odot$ and
$10^{10}$~M$_\odot$.  With a halo mass independent escape fraction for
ionizing photons, this model was able to match constraints from
measurements of the Ly$\alpha$ opacity and the Thomson scattering
optical depth.

The blue curves in Figure~\ref{fig:ps_nonstd} show the 21~cm power
spectrum in our simulation at $z=7$, $8$, and $10$ when the number of
ionizing photons emitted by a halo is assumed to scale with halo mass
as $M_\mathrm{halo}^{1.41}$.  Table~\ref{tab:models} lists other
properties of this model and a slice through the 21~cm brightness
field in this model is shown in Appendix~\ref{sec:2models}.  The model
is calibrated to the Late/Default reionization history.  Overall, the
results are not much different from the Late/Default model presented
earlier.  The power increases throughout the range of $k$ values
shown, by about 20\%, because of the additional clustering added to
the ionization field by the superlinear halo mass scaling of the
number of ionizing photons.

\subsection{Effect of feedback}
\label{sec:fback}

Strong photoionization feedback can prevent star formation in low mass
haloes due to photoevaporation of gas \citep{1998MNRAS.296...44G,
  2008MNRAS.390..920O, 2009MNRAS.399..369N, 2016arXiv160307729G}.
Feedback from supernovae is likely even more effective and seems to be
a critical ingredient in understanding the abundance of faint galaxies
(e.g., \citealt{2003ApJ...599...38B}, \citealt{2013MNRAS.428.2966P}).
If feedback is strong in low mass galaxies, the ionizing photon
emissivity is dominated by high mass haloes.  This can have a
significant effect on the topology of the ionized regions
\citep{2012MNRAS.423.2222I, 2015arXiv151200564G}.  To consider how
this affects our prediction for the 21~cm power spectrum, in
Figure~\ref{fig:ps_nonstd} we consider a model in which only high mass
haloes are present (red curves).  The minimum halo mass considered in
this model is $3.5\times 10^{10}$~M$_\odot$ which is a factor 152
higher than the minimum halo mass considered in the Late/Default model
above.  The model is calibrated to the Late/Default reionization
history.  Table~\ref{tab:models} gives further details.

Our results for the effect of very strong feedback on the 21~cm power
spectrum are in agreement with previous findings in the literature
\citep{2007MNRAS.377.1043M, 2012MNRAS.423.2222I}.  As seen in
Figure~\ref{fig:ps_nonstd}, due to the enhanced clustering of high
mass sources the 21~cm power is enhanced across all scales by factors
of a few.  For a given volume-weighted ionization fraction $Q_V$ the
ionization field is now dominated by few large ionized regions.  These
regions are also more clustered, which increases the 21~cm power.  A
slice through the 21~cm brightness field in this model is shown in
Appendix~\ref{sec:2models}.  Although broad features in the power
spectrum are qualitatively similar to those in the Late/Default model,
at redshift $z=10$ the large scale power is enhanced almost ten times
to about $100$ mK$^2$.  Thus reionization histories with strongly
clustered bright sources are favourable for the detection of the 21~cm
signal.  Recently, it has been argued that faint active galactic
nuclei (AGN) could contribute significantly to the ionizing background
at $z\sim 6$ \citep{2015A&A...578A..83G, 2015ApJ...813L...8M,
  2016arXiv160608231C}.  Such a scenario would result in an
enhancement in the large-scale 21~cm power.

\begin{figure*}
\begin{center}
  \begin{tabular}{cc}
    \includegraphics*[width=\columnwidth]{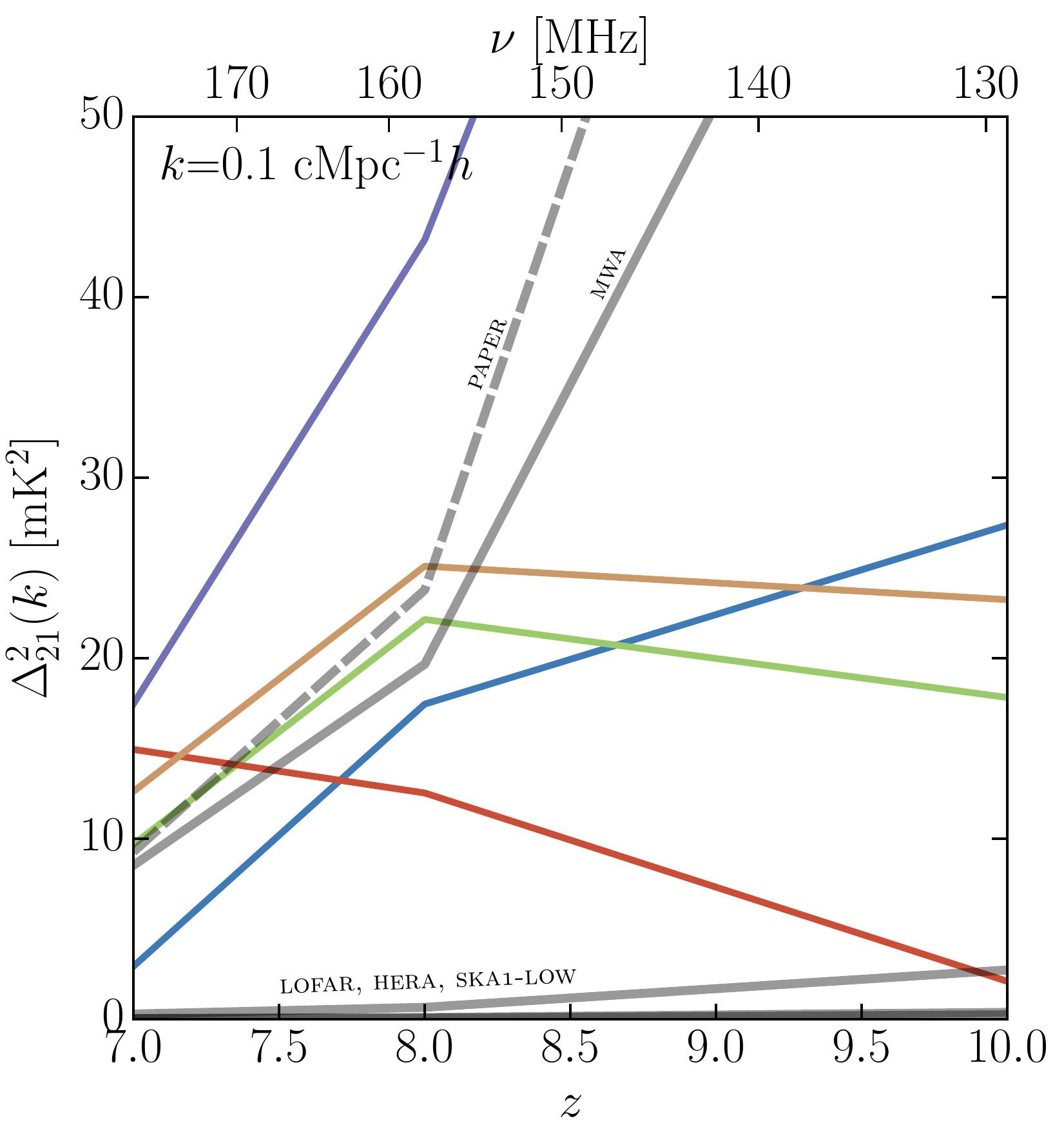} &
    \includegraphics*[width=\columnwidth]{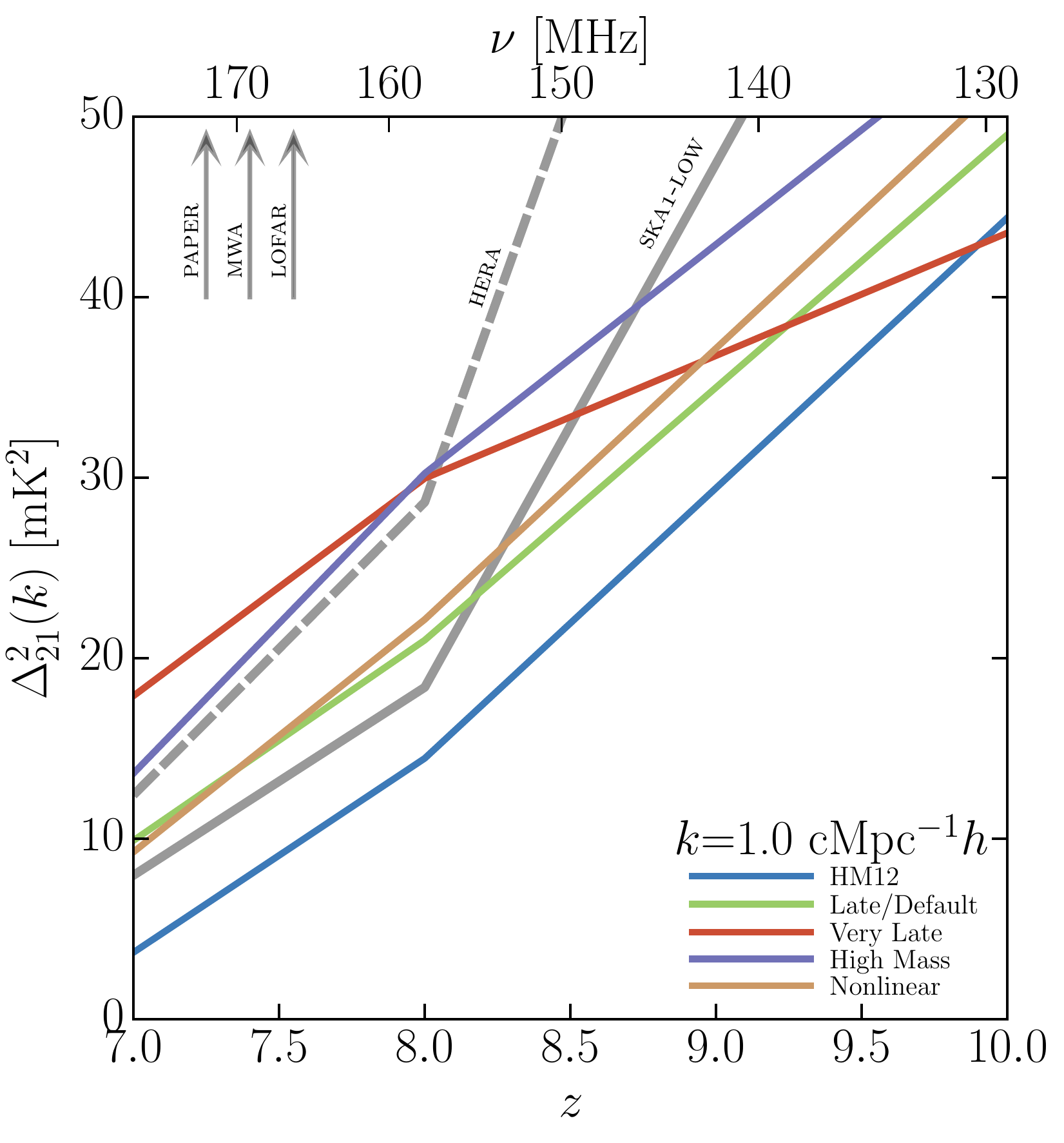} \\
  \end{tabular}
\end{center}
\caption{Evolution of 21~cm power at $k=0.1$ cMpc$^{-1}h$ (left panel)
  and $k=1.0$ cMpc$^{-1}h$ (right panel) in all five models presented
  in this paper: HM12 (blue), Late/Default (green), Very Late (red),
  High Mass (orange), and Nonlinear (beige).  Grey curves show
  experimental sensitivities at the two scales. At $k=0.1$
  cMpc$^{-1}h$, LOFAR, HERA, and SKA1-LOW, have much lower thermal
  noise than the signal.  At $k=0.1$ cMpc$^{-1}h$, PAPER, MWA, and
  LOFAR, have much higher noise than the signal.}
\label{fig:ke}
\end{figure*}

\subsection{Evolution of power spectra}

Figure~\ref{fig:ke} shows the evolution of 21~cm power from $z=10$ to
$z=7$ in all of the reionization models considered in this paper.  The
left panel shows the evolution of power at $k=0.1$ cMpc$^{-1}h$ while
the right panel shows the power at $k=1.0$ cMpc$^{-1}h$.  The blue,
green, and red curves show results from the HM12, Late/Default, and
Very Late models respectively.  Orange curves show results from the
High Mass model calibrated to the Late/Default reionization history.
Beige curves shows the Nonlinear model which is also calibrated to the
Late/Default reionization history.

We find that the small scale power ($k=1.0$ cMpc$^{-1}h$) decreases
with redshift in all five of our models.  The power at these scales is
mostly affected by the reducing ionization fraction and
photoionization rates.  All three models calibrated to the
Late/Default reionization history have very similar behaviour at this
scale.  The High Mass model shows higher amplitude at all redshifts
due to the higher clustering of ionizing sources in this model.  The
HM12 and Very Late models behave differently.  The HM12 model has
lower power due to smaller neutral fraction, and the Very Late model
has higher power.

There is a larger qualitative variation in the behaviour of different
models at $k=0.1$ cMpc$^{-1}h$.  The power at this large scale is
related to the the variance of the 21~cm brightness temperature
distribution that could be measured by the 21~cm experiments
\citep{2014MNRAS.443.1113P, 2014MNRAS.443.3090W}.  The Late/Default
model, which agrees with most other reionization constraints, shows
the rise and fall signature in the power at this scale.  It should be
possible for most experiments to detect this.  In the redshift range
$z=7$--$10$ considered here, the HM12 model only shows the falling
part of the power evolution, while the Very Late model only shows the
rising part.  The High Mass model shows a very rapid increase in power
at this scale towards high redshift.  As discussed above, for the same
volume-weighted ionized fraction the ionization field in this model is
dominated by a few large regions around high mass sources.  These
sources are highly clustered, which correspondingly increases the
21~cm power.  The Nonlinear and Late/Default models have similar
behaviour at this scale.

\section{Conclusions}
\label{sec:end}

We have presented here predictions of the spatial distribution of the
21~cm brightness temperature from the epoch of reionization based on
high dynamic range cosmological hydrodynamical simulations from the
Sherwood simulation suite \citep{2016arXiv160503462B} for reionization
histories motivated by constraints from \lya absorption and emission
data as well as CMB data.  Our models of the 21~cm signal were
obtained by combining the high dynamic range cosmological simulations
with excursion set based models of the growth of ionized regions
during reionization.  This has allowed us to efficiently obtain
realistic 21~cm predictions that are firmly anchored in current
constraints on how reionization ends and that have sufficient
resolution to account for the self-shielding of neutral hydrogen in
dense regions within otherwise fully ionized regions.  Our main
conclusions are as follows:

\begin{itemize}
  \item Our preferred `Late/Default' model of reionization is
    consistent with a variety of observational constraints such as the
    electron scattering optical depth, Ly$\alpha$ opacity, galaxy UV
    luminosity function at high redshifts, and estimates of the
    hydrogen photoionization rates from quasar absorption spectra
    \citep{2015MNRAS.453.2943C}.  In this model the volume-weighted
    ionization fraction $Q_V$ evolves from 0.37 at $z=10$ to 0.82 at
    $z=7$ and reionization is complete at redshift $z\sim 6$.  The
    variance of the 21~cm brightness temperature field at large scales
    accessible to current and upcoming experiments ($k\sim 0.1$
    cMpc$^{-1}h$) reduces from about 20 mK$^2$ at $z=10$ to about 10
    mK$^2$ at $z=7$.  Power at these large scales first increases from
    $z=10$ to $8$, when $Q_V$ is 0.65, and then decreases down to
    $z=7$.  The change from the rise to the fall of the signal of
    reionization, where the 21~cm power peaks occurs when $Q_V$ is
    about 50\%, which occurs at about $z=8.5$ in this model.  The
    small scale 21~cm power in this model decreases continuously from
    about 50 mK$^2$ at $z=10$ to 10 mK$^2$ at $z=7$ ($k\sim 1$
    cMpc$^{-1}h$).  These small scales are, however, unlikely to be
    accessible to any of the five experiments that we have
    investigated here, particularly at $z=10$.

  \item Self-shielding in high density regions within ionized regions
    affects the 21~cm power in two ways.  At the large scales
    accessible to experiments, self-shielding decreases power by up to
    30\%.  The contribution to the 21~cm power from self-shielded
    regions tends to be greater at high redshifts unless the
    ionization fractions are too small.  At small scales,
    self-shielding enhances the 21~cm power by adding structure within
    ionized regions.  This effect is generally stronger at lower
    redshift due to the larger volume occupied by ionized regions.
    The enhancement in power at small scales due to self-shielding can
    be considerable, often even greater than an order of
    magnitude. Unfortunately, these scales are too small to be
    detected by the 21~cm experiments considered here.  By suppressing
    power at intermediate scales self-shielding can reduce the rise
    and fall signature of the epoch of reionization.

  \item In addition to our favoured Late/Default model, we have
    considered two other reionization histories.  In the reionization
    history predicted by the widely used HM12 model of the
    meta-galactic UV background, reionization occurs earlier and the
    21~cm signal peaks as early as $z=10$.  Otherwise the evolution of
    the 21~cm brightness distribution is qualitatively similar to the
    Late/Default model on all scales.  As a result of the earlier rise
    of the ionized volume fraction the large scale power does not show
    the same rise and fall behaviour in this model between $z=10$ and
    $7$ due to the relatively larger ionization fraction already at
    $z=10$.  Instead the large scale variance of the 21~cm brightness
    decreases continuously from about 30 mK$^2$ at $z=10$ to about 3
    mK$^2$ at $z=7$ in this model.  Our third reionization history,
    the `Very Late' model, in which reionization also ends at $z=6$
    but starts later than in the other models, also does not show the
    rise and fall behaviour in the large scale 21~cm power.  In this
    model the large scale power increases continuously from about 2
    mK$^2$ at $z=10$ to about 15 mK$^2$ at $z=7$. The effect of
    self-shielding on the power spectrum in these two models is
    qualitatively similar to that in the Late model. In all three
    cases the effect of self-shielding reduces the large scale 21~cm
    power by up to 30\% and increases the small scale power by factors
    of up to 10.  The latter effect is not accessible at the
    resolution of the 21~cm experiments we have considered here.

  \item We have also varied the scaling of the luminosity of the
    ionizing sources with host halo mass in our models. Our
    `Nonlinear' model has the same reionization history as the
    Late/Default model but an ionizing photon rate that is assumed to
    be a nonlinear function of the halo mass, as, e.g., preferred by
    the radiation hydrodynamical simulations of
    \citet{2011MNRAS.410.1703F}.  This moderately strengthens the
    relative role of high mass haloes compared to low mass haloes
    during reionization.  The results of this model are nearly
    identical to those from our Late/Default model.  We have further
    considered a higher cut-off in the mass of haloes hosting ionizing
    sources.  In this `High Mass' model, we remove low-mass haloes
    from our ionization field calculation.  This scenario corresponds
    to strong feedback, completely suppressing star formation in
    low-mass haloes.  We find that although this model is calibrated
    to the same reionization history as our Late/Default model,
    suppression of low luminosity sources changes the 21~cm signal
    significantly.  Broad features in the power spectrum are
    qualitatively similar to those in the Late/Default model, but at
    redshift $z=10$ the large scale power is enhanced almost ten times
    to about $100$ mK$^2$.  Thus reionization histories with strongly
    clustered bright sources would clearly favour the detection of the
    21~cm signal.

  \item We have derived sensitivity limits of five current and
    upcoming experiments, PAPER, MWA, LOFAR, HERA, and SKA1-LOW, for
    the 21~cm power spectrum due to thermal noise and consider the
    prospects of detecting the reduced large scale 21~cm power
    spectrum in presence of self-shielding.  Assuming perfect
    foreground removal, LOFAR, HERA, and SKA1-LOW should be able to
    detect the power spectrum at $k\sim 0.1$ cMpc$^{-1}h$ at $z=8$ at
    20-$\sigma$ to 100-$\sigma$ significance in our Late/Default
    model, assuming perfect foreground removal.  The significance
    drops at $z=7$ as well as $z=10$ due to changes in the ionization
    structure and evolution in experimental sensitivities.  Detection
    is more difficult with the first generation experiments PAPER and
    MWA except at scales larger than our box size of 160 $h^{-1}$cMpc,
    although foreground removal can be difficult at these large scales
    \citep{2014ApJ...782...66P}.  At $k=0.1$ cMpc$^{-1}h$, MWA should
    be able to detect the rise and fall signal of the epoch of
    reionization at less than 2-$\sigma$, whereas LOFAR should be able
    to detect it at nearly 10-$\sigma$.  Assuming ideal foreground
    removal, HERA and SKA1-LOW should be able to detect this signature
    comfortably at excess of 50-$\sigma$. At their design sensitivity
    LOFAR, HERA and SKA1-LOW should therefore easily discriminate
    between the ionization histories presented here.
    
\end{itemize}

The calibration procedure used in this paper provides a relatively
low-cost method of performing high dynamic range simulations of the
cosmological 21~cm signal for reionization histories that are well
anchored in constraints from other data on how reionization ends.
Although self-consistent large scale simulations of cosmic
reionization are now gradually becoming possible, the method presented
in this paper is valuable for an efficient and flexible exploration of
the relevant parameter space that will be necessary for inference from
the statistical detection of the 21~cm signal.

\section*{Acknowledgments}

We thank the referee for helpful comments and also acknowledge useful
discussions with Jonathan Chardin, Kanan Datta, Raghunath Ghara,
Geraint Harker, Joseph Hennawi, Harley Katz, Sergey Koposov, Jonathan
Pritchard, Sijing Shen, and Saleem Zaroubi.  Support by ERC Advanced
Grant 320596 `The Emergence of Structure During the Epoch of
Reionization' is gratefully acknowledged.  EP gratefully acknowledges
support by the Kavli Foundation.  We acknowledge PRACE for awarding us
access to the Curie supercomputer, based in France at the Tr\'es Grand
Centre de Calcul (TGCC).  This work used the DiRAC Data Centric system
at Durham University, operated by the Institute for Computational
Cosmology on behalf of the STFC DiRAC HPC Facility
(www.dirac.ac.uk). This equipment was funded by BIS National
E-infrastructure capital grant ST/K00042X/1, STFC capital grants
ST/H008519/1 and ST/K00087X/1, STFC DiRAC Operations grant
ST/K003267/1 and Durham University. DiRAC is part of the National
E-Infrastructure.  This research was supported by the Munich Institute
for Astro- and Particle Physics (MIAPP) of the DFG cluster of
excellence ``Origin and Structure of the Universe''.

\bibliographystyle{mnras}
\bibliography{refs}

\appendix

\section{Convergence test}
\label{sec:conv}

We have discussed the role of self-shielded structure in the IGM on
the large scale ionization field and the 21~cm signal.  The base
cosmological simulation that we used for our discussion has a box size
of 160 $h^{-1}$cMpc and $2048^3$ gas and dark matter particles.  This
corresponds to a dark matter particle mass of
$M_\mathrm{dm}=3.44\times 10^7$ $h^{-1}$M$_\odot$ and gas particle
mass of $M_\mathrm{gas}=6.38\times 10^6$ $h^{-1}$M$_\odot$.  It is
because of this high mass resolution that the simulation was able to
resolve the self-shielded structure that is usually missed by
radiative transfer simulations of the epoch of reionization.  We now
check the convergence of our results by deriving the 21~cm signal in a
cosmological simulation with higher resolution than our base
simulation.

\begin{figure}
  \begin{center}
    \includegraphics*[width=\columnwidth]{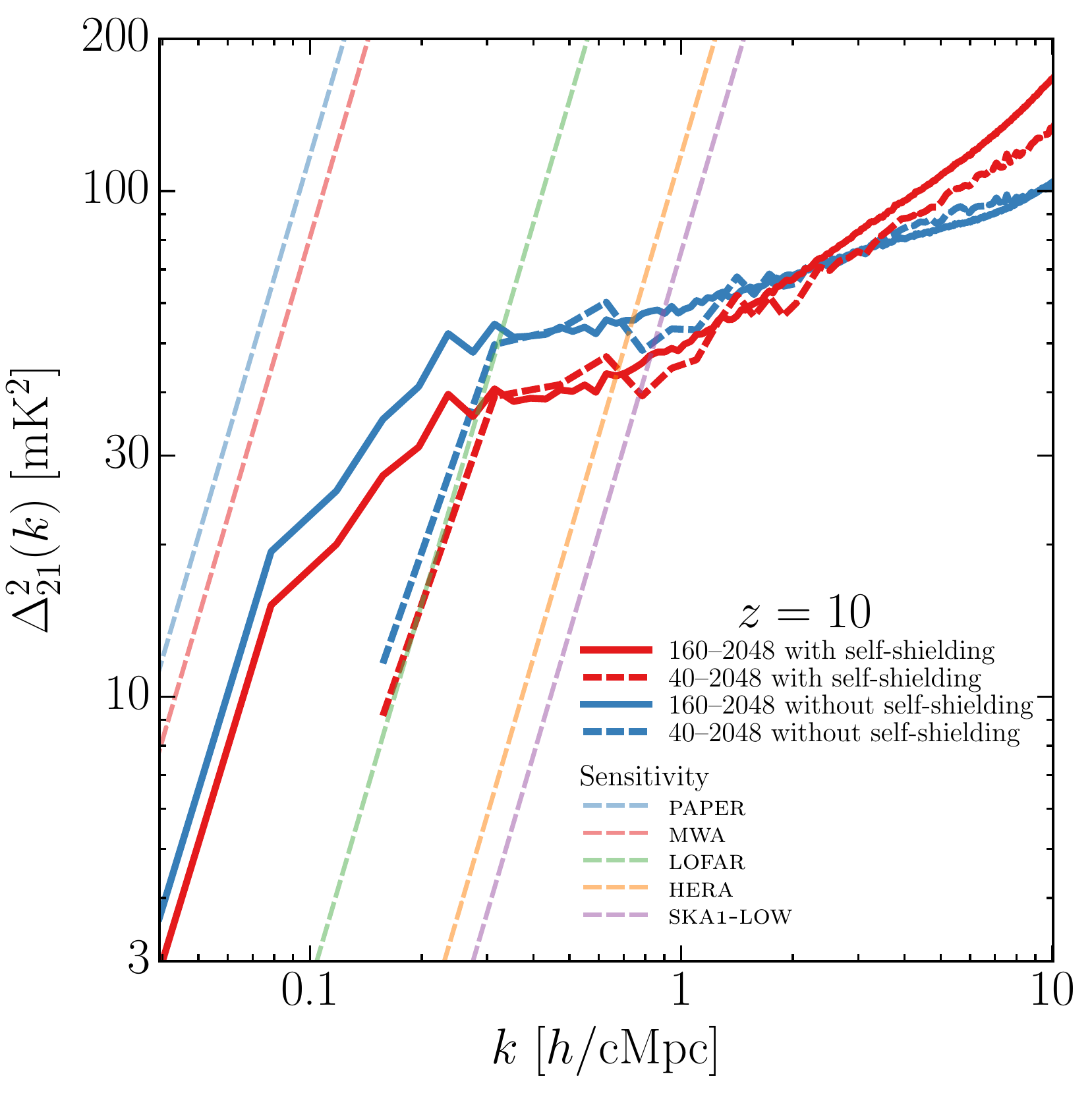}
  \end{center}
  \caption{Effect of self-shielding on the 21~cm power spectrum at
    $z=10$ at two different numerical resolutions.  The solid red and
    blue curves show the power spectra with and without
    self-shielding, respectively, in our fiducial base simulation,
    which has a box size of 160 $h^{-1}$cMpc and 2048$^3$ gas
    particles.  The dashed red and blue curves show the power spectra
    from a simulation with 64 times higher mass resolution.  It has a
    box size of 40 $h^{-1}$cMpc and 2048$^3$ gas particles.  Identical
    reionization histories and halo mass ranges are used to derive
    these power spectra.  The faint dashed curves show experimental
    sensitivities.  Power spectra from the two simulations agree well,
    suggesting numerical convergence.}
\label{fig:resolution_study}
\end{figure}

\begin{figure*}
  \begin{center}
  \includegraphics*[width=\textwidth]{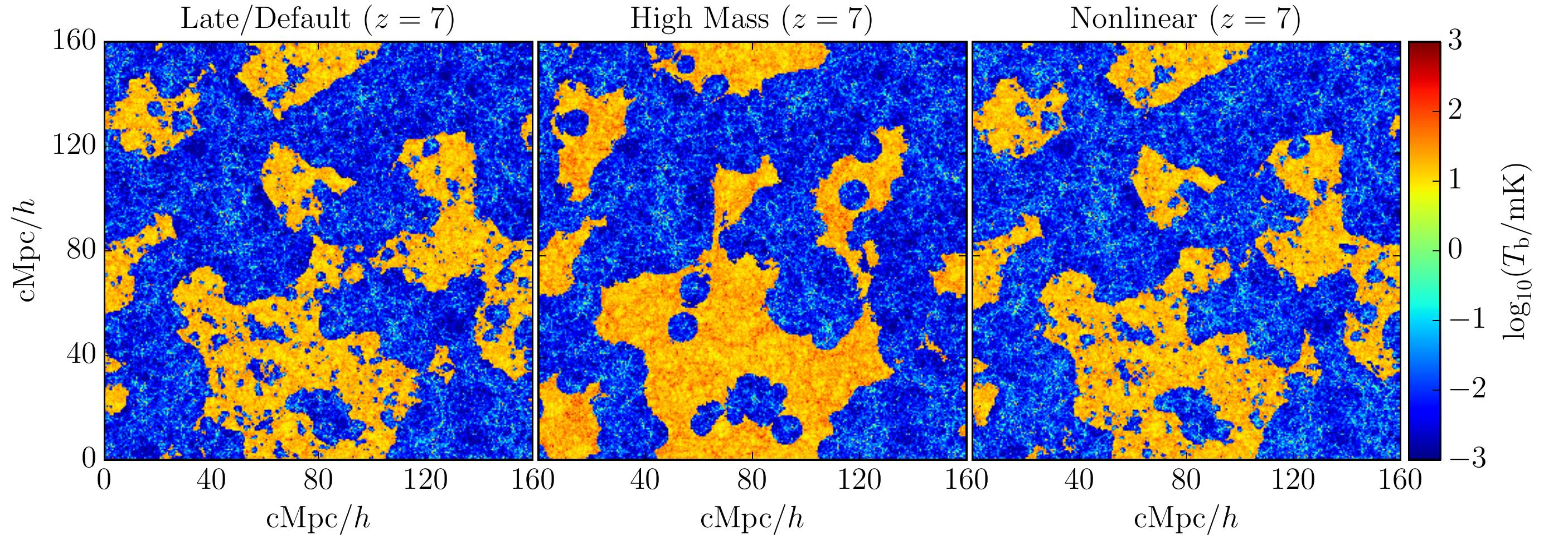}
\end{center}
\caption{Distribution of 21~cm brightness in the Late/Default (left
  panel), High Mass (middle panel), and Nonlinear (right panel) models
  at $z=7$.  Each slice has a depth of 78.1 $h^{-1}$ckpc.  All three
  simulations are calibrated to the Late/Default reionization history,
  which has $Q_V=0.82$ at this redshift.  Other details of these three
  models are given in Table~\ref{tab:models}.}
\label{fig:tb3}
\end{figure*}

Figure~\ref{fig:resolution_study} shows the 21~cm power spectrum at
$z=10$ taken from a simulation with a box size of 40 $h^{-1}$cMpc and
$2048^3$ gas and dark matter particles.  This corresponds to a dark
matter particle mass of $M_\mathrm{dm}=5.37\times 10^5$
$h^{-1}$M$_\odot$ and gas particle mass of $M_\mathrm{gas}=9.97\times
10^4$ $h^{-1}$M$_\odot$.  This simulation thus has a factor of 64
higher mass resolution than our base simulation.  The spatial
resolution is also higher as the softening length is set to
$l_\mathrm{soft}=0.78$ $h^{-1}$ckpc.  This simulation is also a part
of the Sherwood suite of simulations \citep{2016arXiv160503462B}.
Other details of this simulation are identical to our base simulation.
The minimum halo mass in our base simulation is $1.6\times 10^{8}
h^{-1}$M$_\odot$; the maximum halo mass is $2.1\times 10^{12}
h^{-1}$M$_\odot$ at $z=7$.  In comparison, the minimum halo mass in
the higher resolution simulation is $2.4\times 10^{6}
h^{-1}$M$_\odot$.  The maximum halo mass is $4.2\times 10^{10}
h^{-1}$~M$_\odot$ at $z=7$.  For Figure~\ref{fig:resolution_study}, a
minimum halo mass of $1.6\times 10^{8} h^{-1}$M$_\odot$ is used when
placing the ionizing sources so that the large scale ionization fields
can be compared.

In Figure~\ref{fig:resolution_study}, the solid red and blue curves
show the 21~cm brightness temperature power spectra in the low
resolution simulation with and without the effect of self-shielding
respectively.  The dashed curves show the corresponding power spectra
from the high resolution simulation.  Self-shielding is implemented
using the same prescription as in Section~\ref{sec:ss}.  Both
simulations are calibrated to the Late/Default model of reionization
described in Section~\ref{sec:calibration}.  The power spectrum curves
of the low resolution simulation are identical to those from
Figure~\ref{fig:ps_3zs}.  We find that power spectra from the two
simulations agree quite well at scales relevant to experiments
($k<1$~cMpc$^{-1}h$).  The effect of self-shielding is also identical
in the two simulations on observationally accessible scales.  As
described in Section~\ref{sec:ss_effect} self-shielding reduces the
large scale power by about a factor of two.  This reduction is
identical in the two simulations, showing that our self-shielding
implementation has converged in the base simulation.

\section{21~cm distribution in the High Mass and Nonlinear models}
\label{sec:2models}

Figure~\ref{fig:tb3} shows the 21~cm brightness distribution at $z=7$
in the High Mass and Nonlinear models that are discussed in Sections
\ref{sec:nonlin} and \ref{sec:fback}.  The Late/Default reionization
model is shown in the left panel for comparison.  The middle panel
shows the High Mass model in which sources are placed only in high
mass haloes.  The right panel shows the Nonlinear model in which the
source emissivity has a nonlinear dependence on the halo mass.  All
three models are calibrated to the Late/Default reionization history
so that they have $Q_V=0.82$ at this redshift.

We find that the 21~cm field in the fiducial Late/Default model is
nearly identical to that in the Nonlinear model.  This is
understandable as the nonlinearity implemented here is mild
($N_\mathrm{gamma} \propto M_\mathrm{halo}^{1.41}$).  The High Mass
case shows larger differences from the fiducial Late/Default model.
Smaller ionized regions around small mass haloes are not present in
the High Mass model.  These differences in the ionization fields are
reflected in the 21~cm power spectra for these models, which are shown
in Figure~\ref{fig:ps_nonstd}.

\bsp
\label{lastpage}
\end{document}